\documentclass[a4paper,onecolumn,11pt,accepted=2024-10-22]{quantumarticle}
\pdfoutput=1

\usepackage[numbers,sort&compress]{natbib}

\usepackage{amsmath}

\usepackage{comment}
\usepackage{graphicx}%
\usepackage{bm}%
\usepackage[colorlinks = true,
            linkcolor = blue,
            urlcolor  = blue,
            citecolor = blue,
            anchorcolor = blue]{hyperref}
\usepackage{physics}
\usepackage{soul}
\usepackage{mathtools}

\usepackage{multirow}%
\usepackage{amsmath,amssymb,amsfonts}%
\usepackage{amsthm}%
\usepackage{mathrsfs}%
\usepackage[title]{appendix}%
\usepackage{xcolor}%
\usepackage{textcomp}%
\usepackage{booktabs}%
\usepackage{algorithm}%
\usepackage{algorithmicx}%
\usepackage{algpseudocode}%
\usepackage{listings}%

\usepackage[scr=boondoxo]{mathalfa}	%

\usepackage{braket}         %
\usepackage{colonequals}    %
\usepackage{tensor}		    %
\usepackage{extpfeil}		%

\newcommand{\lb}{\left}	    %
\newcommand{\rb}{\right}    %

\newcommand{\tx}{\text}						%
\newcommand{\bd}{\boldsymbol}				%
\newcommand{\mc}{\mathcal}					%
\newcommand{\ms}{\mathscr}					%
\newcommand{\qew}{\mathscr{q}}				%
\newcommand{\bb}{\mathbb}					%

\newcommand{\nn}{\nonumber}				     	%
\newcommand{\qd}{\qquad}				    	%
\newcommand{\qqd}{\qquad\qquad}				    %
\newcommand{\qqqd}{\qquad\qquad\qquad}		    %
\newcommand{\qqqqd}{\qquad\qquad\qquad\qquad}	%

\newcommand{\infi}{\infty}      %
\newcommand{\ct}{\dagger}       %
\newcommand{\trps}{\intercal}   %
\newcommand{\ce}{\colonequals}	%
\newcommand{\imp}{\implies}		%
\newcommand{\ti}{\widetilde}	%
\newcommand{\oa}{\hat{a}}	    %
\newcommand{\oad}{\hat{a}^{\dagger}}    %

\newcommand{\iu}{\mathrm{i}}    %
\newcommand{\ec}{\mathrm{e}}    %
\newcommand{\ddd}{\mathrm{d}}	%

\newcommand{\Abs}[1]{\left|{#1}\right|}		%
\newcommand{\Bk}{\Braket}                   %
\newcommand{\Kb}[2]{\Ket{#1}\Bra{#2}}		%

\newcommand{\intall}{\int_{-\infty}^{\infty}}   %
\newcommand{\intallr}{\int_{0}^{\infty}}        %

\newcommand{\txx}{\text{x}} %
\newcommand{\txy}{\text{y}} %

\newcommand{\hypg}[3]{\tensor[_#1]{#2}{_#3}}	%

\newcommand{\eqn}[1]{\begin{equation}#1\end{equation}}				%
\newcommand{\eqns}[1]{\begin{equation*}#1\end{equation*}}			%
\newcommand{\aln}[1]{\begin{align}#1\end{align}}					%
\newcommand{\alns}[1]{\begin{align*}#1\end{align*}}					%
\newcommand{\subeqn}[1]{\begin{subequations}#1\end{subequations}}	%
\newcommand{\pmx}[1]{\begin{pmatrix}#1\end{pmatrix}}	            %
\newcommand{\pw}[1]{\begin{cases}#1\end{cases}}		                %

\usepackage[section]{placeins}  %
\newcommand{\FB}{\FloatBarrier} %

\makeatletter
	\newcommand{\vast}{\bBigg@{3}}
	\newcommand{\Vast}{\bBigg@{4}}
\makeatother

\renewcommand{\Re}{\operatorname{Re}}
\renewcommand{\Im}{\operatorname{Im}}

\newtheorem{remark}{Remark}

\usepackage{todonotes}

\usepackage[numbers]{natbib}

\begin{document}

\title{Quantum kernel machine learning with continuous variables}

\author{Laura J. Henderson}
\affiliation{School of Mathematics and Physics, The University of Queensland, QLD 4072, Australia}
\affiliation{ARC Centre for Engineered Quantum Systems, The University of Queensland, QLD, 4072, Australia.}
\author{Rishi Goel}
\affiliation{School of Mathematics and Physics, The University of Queensland, QLD 4072, Australia}
\author{Sally Shrapnel}
\affiliation{School of Mathematics and Physics, The University of Queensland, QLD 4072, Australia}
\affiliation{ARC Centre for Engineered Quantum Systems, The University of Queensland, QLD, 4072, Australia.}

\begin{abstract}
The popular qubit framework has dominated recent work on quantum kernel machine learning, with results characterising expressivity, learnability and generalisation. As yet, there is no comparative framework to understand these concepts for continuous variable (CV) quantum computing platforms. In this paper we represent CV quantum kernels as closed form functions and use this representation to provide several important theoretical insights. We derive a general closed form solution for all CV quantum kernels and show every such kernel can be expressed as the product of a Gaussian and an algebraic function of the parameters of the feature map. Furthermore, in the multi-mode case, we present quantification of a quantum-classical separation for all quantum kernels via a hierarchical notion of the ``stellar rank" of the quantum kernel feature map.
We then prove kernels defined by feature maps of infinite stellar rank, such as GKP-state encodings, can be approximated arbitrarily well by kernels defined by feature maps of finite stellar rank. Finally, we simulate learning with a single-mode displaced Fock state encoding and show that (i) accuracy on our specific task (an annular data set) increases with stellar rank, (ii) for underfit models, accuracy can be improved by increasing a bandwidth hyperparameter, and (iii) for noisy data that is overfit, decreasing the bandwidth will improve generalisation but does so at the cost of effective stellar rank. 
\end{abstract}

\maketitle

\tableofcontents

\section{Introduction}\label{Introduction Section}

The quantum machine learning (QML) community has recently begun to explore whether quantum resources may be useful for kernel machine learning \cite{schuld2021-quantumkernel,schuld_PhysRevLett.122.040504,Wittek_2016}. While research has typically focused on improving traditional classical kernel methods, such as support vector machines, classical kernelisation is in fact far more ubiquitous. Kernels, which essentially provide a similarity metric between data points, appear as filters in convolutional neural networks \cite{Goodfellow-et-al-2016}, can represent attention matrices in transformer networks \cite{tsai-etal-2019-transformer}, are used as training signals for generative networks \cite{bińkowski2018demystifying}, and can provide a key mechanism for causal discovery \cite{mitrovic2018causal}. Clearly, there is much to be gained by understanding whether quantum kernels can provide an advantage over their classical counterparts \cite{4270130,5337958,akaho2007kernel,pmlr-v80-belkin18a,NIPS2009_5751ec3e}.

This recent exploration has led to the development of quantum kernel selection tools \cite{Hubregtsen_2022}, generalisation bounds \cite{Gyurik_2023}, optimal solution guarantees \cite{schuld2021-quantumkernel,Jerbi_2023}, and has resulted in several physical implementations \cite{Liu_2021,glick2022covariant,Havl_ek_2019,doi:10.1126/science.abn7293}. The community has learned that although entangled quantum kernels—including those generated by deep parametrised quantum neural networks (PQNN)—are highly expressive, such expressivity typically comes at a cost. This is the so-called ``exponential concentration" problem analogous to the barren plateau problem in quantum neural networks \cite{McClean_2018} —as quantum kernels become more expressive, they typically also become exponentially harder to learn and less likely to generalise \cite{thanasilp2022exponential, kübler2021inductive, Huang_2021}. Essentially, the value of the kernel between different datapoints decreases as a function of the size of the problem—for discrete variable quantum kernels, a highly expressive kernel yields exponential concentration. Recent numerical work suggests it may nonetheless be possible to overcome these learning difficulties by manipulating a bandwidth hyper-parameter to tune the expressivity of the quantum kernel \cite{canatar2023bandwidth,canatarthesis}, a technique inspired by bandwidth tuning of classical Gaussian kernels \cite{Schölkopf_Smola_2002}. Finding the sweet spot where the quantum kernel is both learnable and generalises well, but is nonetheless still classically hard to simulate is, however, an open challenge \cite{Slattery_2023}. Robustness to noise also presents a further unexplored challenge to such kernel tuning techniques.

As a consequence, the QML community has to some extent converged on a new quest. Rather than seeking a quantum advantage for kernel machine learning \textit{per se,} physicists are now searching for inductive biases that specific quantum kernels, or families of quantum kernels, may bring to particular ML tasks \cite{kübler2021inductive}. The thinking is inspired by the tremendous advantage convolutional neural networks have provided imaging tasks due to their translational invariance \cite{Goodfellow-et-al-2016}. To this end, the group theoretic structure of some specific quantum kernels has been used to exploit structure in certain classical learning problems to prove quantum advantage \cite{glick2022covariant,Liu_2021, ragone2023representation}. As such, there is strong motivation to identify and understand new classes of quantum kernels.

An outstanding key challenge to substantial progress is the theoretically opaque nature of quantum kernels. Classical kernels employ the ``kernel trick"— one avoids explicitly evaluating the kernel in feature space by instead using an analytic representation acting in the original data space (e.g.\ the Gaussian function, or other RBF kernel). This provides direct access to the kernel and permits theoretical analysis. In contrast, quantum kernel values are accessed via estimating inner products through quantum measurement. Thus, quantum measurement expectation values approximate each kernel matrix entry up to some additive error. Functional forms of the kernel are rarely available and theoretical understanding of quantum kernels is somewhat limited.

Interestingly, almost all the work on quantum kernels to date has focused on discrete, finite dimensional quantum systems, such as those generated by parameterised quantum circuits. While a few works have evaluated specific, continuous variable, infinite dimensional quantum encodings \cite{Ghobadi_PhysRevA.104.052403,schuld_PhysRevLett.122.040504, Tiwari_2022_coherent, Liu_2023, Bowie_2023}, there is as yet no unifying approach to CV quantum kernels.

In this paper, we use the holomorphic representation of continuous variable quantum states \cite{Chabaud_2022} to find a closed form expression for an arbitrary CV multi-mode kernel,
which allows for insight into the general theoretical structure of CV quantum kernels—every kernel can be expressed as the product of a Gaussian and an algebraic function of the four key parameters of the feature map. 
This structure permits some preliminary intuition into possible trade-offs between bandwidth hyperparameter tuning (to improve generalisation and learnability) and consequent loss of quantum advantage \cite{Slattery_2023} as kernel values will be close to zero beyond a certain distance.
Furthermore, the use of holomorphic functions allows us to present a framework with a very natural taxonomy of kernel ``quantumness", achieved via the notion of the stellar rank of the quantum kernel feature map. Stellar rank ultimately provides useful guarantees as to the hardness of classical simulation \cite{Chabaud_2022, Chabaud_GcoreComplexity,Chabaud_2023} of such kernels.

The paper is organised as follows: in section \ref{preliminaries section} we introduce relevant mathematical notation and background. In section \ref{Bosonic quantum computing section} we review holomorphic representations of quantum CV systems (based on \cite{Chabaud_2022}). In section \ref{CV Quantum Feature Maps section} we formalise how one may define a CV quantum kernel using the tools of holomorphic functions, show it satisfies necessary properties and comment on the dependence of classical simulability on stellar rank. In section \ref{General Kernel Section}, we present the multi-mode general CV encoding and show that all finite rank kernels can be expressed as the product of a Gaussian and algebraic function term. We also prove that quantum kernels defined by feature maps of infinite stellar rank, such as GKP and cat-state encodings, can be approximated to arbitrary precision with kernels defined by feature maps of finite stellar rank. Section \ref{Examples Section} presents an example of a single mode displaced Fock state encoding.  We use this example kernel to simulate five learning experiments to verify the expected general behaviour, using a uniform data set and a data set with annular structure. General qudit kernels are presented in section \ref{Qudit Kernel Section}, where we show that these kernels are a subset of the general multi-mode case.
While for particular quantum kernels there are undoubtedly simpler analytic forms, our goal here is to highlight the general form, characterise the stellar rank, and provide insight into some general features that are likely to be applicable to all CV quantum kernels. To this end, we explore the notion of bandwidth tuning and how it affects our example kernels. 
We conclude with section \ref{Conclusions and future work section}, where we discuss findings and make suggestions for future work.

\section{Preliminaries}\label{preliminaries section}

\subsection{Notation}\label{notation section}

Here we fix notation and formalise the necessary mathematics.

Vectors are denoted in bold unless otherwise specified (e.g. $\boldsymbol{x}$) as are matrices, the latter with capital letters only (e.g. $\boldsymbol{X}$). Sets and vector spaces are written in mathematical calligraphic font (e.g. $\mathcal{X}$). Complex numbers will be stated explicitly or as 2-dimensional real vectors, most commonly we use $z$. Conjugates of complex vectors or matrices are written with superscript asterisk (e.g. $\boldsymbol{x}^*$). Overlines instead represent completion of sets, e.g.  $\overline{\mathcal{X}}$.

$\Ket{\psi}$ will always be a pure quantum state. We reserve $n$ for the Fock state number of such a quantum state. We write all kernels using $k$, whether they are quantum or classical will be made explicit within the text. We define $\bb{N}_0$ as the set of natural numbers including $0$.

Hypergeometric functions are written as $_i F_j$, with $i,j$ representing the specific form. Polynomials of $x$ are written as $P(x)$ and Gaussians as $G(x)$. As usual, $\Gamma$ is the gamma function:
\eqn{
    \Gamma(z) = \intallr \ddd t\ t^{z-1} \ec^{-t}
}
for $\Re{(z)}>0$. %
The inner product of a specific Hilbert space, $\mathcal{H}$, is written as $\braket{\boldsymbol{x}|\boldsymbol{x}'}_{\mathcal{H}}$. Unless otherwise specified, the norm $\norm{\boldsymbol{x}}_\mathcal{H}$ is given by $\sqrt{\braket{\boldsymbol{x}|\boldsymbol{x}}}_\mathcal{H}$ where $\mathcal{H}$ is the space in which $\boldsymbol{x}$ has a well defined inner product.

We define holomorphic functions, denoted by $F^\star$, as complex functions which are complex differentiable in a neighbourhood about every point. $F^\star_{\txx}$ is a holomorphic function dependent on some classical data $\txx \in \mathcal{X}$. Stellar functions are a subset of holomorphic functions with finite roots and written as the product of a polynomial and Gaussian term. As $n$ is used as our Fock state number, our stellar rank (the number of complex roots of $F^\star$) is also $n$. %

\subsection{Introduction to classical kernel machine learning}\label{Kernel Machine learning section}

The core tenet of kernel machine learning (ML) is the application of linear statistical methods to complex, non-linear data. The data---while not separable in the original data space---can be linearly separated after transformation into a higher dimensional space.  The key advantage from kernel methods is the use of the ‘kernel trick’, where one does not need to explicitly compute the data embedding. This trick has found its way into many applications such as ML classification, regression, and clustering \cite{Schölkopf_Smola_2002}.  

Given some data from the input space, $\mc{X}$, a \textit{kernel} is a function, $k:\mc{X}\times\mc{X} \to \bb{F}$ (where $\bb{F}$ is either the real, $\bb{R}$, or complex numbers, $\bb{C}$), such that for all $\txx,\txx' \in \mc{X}$,
\begin{equation}
    k(\txx,\txx') = \braket{\Phi(\txx)|\Phi(\txx')} _{\mathcal{H}}.
    \label{eq:ClassicalKernel}
\end{equation}
where $\Phi: \mathcal{X} \to \mathcal{H}$ is the feature map, which encodes the data into a Hilbert space, $\mc{H}$.
We usually take the feature map such that the data is not linearly separable in $\mathcal{X}$ but is in $\mathcal{H}$---commonly achieved by taking $\mathcal{H}$ to be a higher dimension than $\mathcal{X}$. %
The kernel, $k$ uses the inner product in our feature space to form a measure of similarity between data.

By Mercer's theorem, any function that is (conjugate) symmetric and
positive semi-definite\footnote{More precisely: $\int k(x,y)f(x)f(y)\ \ddd x\ddd y \ge0 $ for all $f\in L^2$.}, i.e., $\forall \txx_i \in \mathcal{X} $ and any $c_i \in \mathbb{C}^n$
\begin{equation}
    \sum_{i,j} c_i c_j^* k(\txx_i,\txx_j) \geq 0
\end{equation}
is a kernel, and there exits a mapping $\Phi$ such that Eq.\ \eqref{eq:ClassicalKernel} is holds. This is the crux of the kernel trick: the function, $k(\txx,\txx')$, can be computed directly without ever needing to compute $\Phi(\txx)$ or even know what it $\Phi$ is.

For simplicity, we will describe the supervised learning case. %
We are given some labelled data set, $\{(\txx_k,\txy_k), k=1,\dots,M\}$ and aim to find a mapping $f(\txx)$ for new unlabeled data points, where $f(\txx)$ is determined by some structure, pattern or probability distribution within the data. %
The solution to this learning problem is given by, 

\begin{equation}\label{minimised supervised learning problem}
    f^*(\txx) = \underset{h\in \mathcal{H}}{\text{arg min}} \frac{1}{M} \sum_{i=1}^M \mathcal{L}(h(\txx_i),\txy_i) + g( ||h||),
\end{equation}

where we define $\mathcal{L}$ as the loss function characterising the performance of the learned function and $\mathcal{H}$ is the Hilbert space of learning functions we are considering \cite{Hofmann_2008}. $g(\cdot)$ is a monotonically increasing regularisation function to reduce overfitting, favouring a smooth function with better generalisation. 

Associated to every such feature space is a unique Reproducing Kernel Hilbert Space (RKHS). This is a space of functions that can be constructed as the completion of the span of kernels,

\begin{equation}\label{RKHS}
     \mathcal{H}_{\tx{RKHS}} = \overline{\textnormal{span}}\{k(\cdot,\txx_i)|\txx_i\in\mathcal{X}\}.
\end{equation}
where $k$ has the reproducing property:
\eqn{
    \forall\txx\in\mc{X},\ \forall f\in\mc{H}_{\tx{RKHS}},\ f(\txx) = \Bk{f,k(\cdot,\txx)}_{\mc{H}_{\tx{RKHS}}}.
    \label{eq:repprop}
}

The construction of the RKHS permits a solution - by the representer theorem - to equation \ref{minimised supervised learning problem} given by,

\begin{equation}\label{optimal kernel function}
    f^*(\txx) = \sum_{\txx_i \in \mathcal{X}} c_i k(\txx,\txx_i), 
\end{equation}
for some $c_i \in \mathbb{R}$ \cite{Schölkopf_Herbrich_Smola_2001}. 

Formal analysis of classical kernel functions allows one to characterise three key quantities:  learnability,  expressivity and generalisation.  Learnability describes how well the optimal kernel as defined in equation \ref{optimal kernel function} can be found as a function of the size of the problem. Expressivity is used to measure the complexity of problems kernels can linearly classify. If a kernel is universal (i.e. perfectly expressive), it can precisely separate any two given sets from a compact metric space of finite training data \cite{Hofmann_2008}. All universal kernels are also characteristic, and can thus be utilised in probabilistic ML applications \cite{JMLR:v12:sriperumbudur11a}. In section \ref{Examples Section} we provide an example of a characteristic CV quantum kernel. %

Generalisation theory aims to assess the quality of the learning scheme. Generalisation bounds provide a measure of how well the kernel - given some finite data set from a distribution $P$ - can be applied to randomly sampled data from $P$ that was not in the initial data set. The boundedness of evaluational functions in any RKHS, $\{f \in RKHS \, : \,  \norm{f} \leq B\}$ yields generalisation bounds given by 

\begin{equation}
    \mathbb{E}_{\txx\in \mathcal{D}} |h^*(\txx) - f| \leq \frac{2B}{\sqrt{M}}. 
\end{equation}

$M$ is the size of the labelled data set and $B$ is a function of the specific kernel and the loss function from equation \ref{minimised supervised learning problem} \cite{shawe-taylor_cristianini_2004}. Such approaches are often defined in terms of VC dimension or fat-shattering dimension \cite{Vladimir2015}, however, these bounds all represent worst case scenarios and have limited practical relevance. %
In practice, bandwidth hyperperameter tuning, which essentially changes the length scale of the kernel, is most often used to improve generalisation. If bandwidth is too small, the kernel will treat most new data points as very far from any training observation, while a bandwidth that is too large creates a kernel that will treat each new data point as nearly equidistant to all training observations. Neither will result in good generalization and clearly bandwidth tuning can have a profound impact on learnability and generalisation. Such tuning is similarly computationally very expensive, although recent techniques utilising Jacobian control have shown some improvements \cite{allerbo2023bandwidth}. %

\subsection{Background on quantum kernel machine learning}\label{Quantum kernels section}

Recently, formal similarities between kernel methods and quantum machine learning (QML) methods have become well established \cite{schuld2021-quantumkernel}. Essentially, QML methods encode data non-linearly into a higher dimensional Hilbert space in which quantum measurement defines a linear decision boundary. For example, in supervised machine learning we can encode our data in the Hilbert space of the quantum system as $\txx\to \ket{\Phi(\txx)}$ and then learn the measurement that optimally separates the data. Typically, this state is prepared with some unitary gate operator $U_{\theta}(\txx)$ that acts on the vacuum state $\ket{0\dots0}$ such that $U_{\theta}(\txx) \ket{0\dots0} = \ket{\Phi(\txx)}$. The kernel function is then defined using the Hilbert-Schmidt inner product as,

\begin{equation}
    k(\txx,\txx') =  \braket{\Phi(\txx)|\Phi(\txx')}_{\mathcal{H}}.
    \label{eq:GeneralQuantumKernel}
\end{equation}

An important subtlety is that in quantum kernels, we no longer use the same ``kernel trick'' as in the classical case; kernel entries are not evaluated using a closed form function, $k(\txx,\txx')$, on the original data. They are instead, approximated via the quantum measurement of $\Abs{\Bk{\Phi(\txx)|\Phi(\txx')}}^2$. This also means that the RKHS of quantum kernels is rarely characterised, though the existence of the quantum feature map directly implies the existence of a RKHS.

While the majority of quantum kernels are characterised using the qubit circuit formalism, several specific examples of CV quantum kernels exist. Schuld's excellent summary paper includes a description of a coherent state kernel encoding \cite{schuld_PhysRevLett.122.040504} and Tiwari et. al. construct a mathematical representation of coherent quantum kernels using generalised hypergeometric functions \cite{Tiwari_2022_coherent}. Ghobadi presents a single-mode squeezed and a single photon (Fock) state quantum kernel and derives a non-classicality witness - a necessary but insufficient condition for quantum advantage - for each \cite{Ghobadi_PhysRevA.104.052403}, and Bowie et. al. describe an experimental platform which exploits Hong–Ou–Mandel interference to evaluate a kernel based on a temporal encoding \cite{Bowie_2023}. There is to date, however, no unifying framework from which to understand these individual results. In the following section we introduce the relevant background on holomorphic representations of CV quantum states to understand our approach.

\section{CV quantum kernels}\label{CV kernels}

\subsection{Representing CV quantum states as holomorphic functions}\label{Bosonic quantum computing section}

Quantum information processing (QIP) is often separated into two paradigms: discrete variable QIP and continuous variable QIP. The former utilises finite dimensional Hilbert spaces and qubits or qudits, whereas the latter utilises infinite dimensional Hilbert spaces and qumodes. In the discrete case, non-Clifford operations or magic states are identified as necessary for bounded error quantum polynomial (BQP) complete, non-classically simulable, computation \cite{gottesman}. Analogously, non-Gaussian operations or non-Gaussian states are identified as necessary for BQP-complete computation in the CV setting \cite{bartlett}. In recent work, Chabaud et. al. present a measure of non-Gaussianity which permits a more nuanced quantification of the computational power of CV quantum computing platforms \cite{Chabaud_2022}. In this approach, CV states are fully characterised by holomorphic functions. Such functions can be thought of as quasi-probability distributions, similar to Wigner functions or Husimi functions. %
In the CV case, we can decompose a finite rank holomorphic function as a stellar function - a product of a Gaussian and polynomial in $z$. The polynomial is characterised by its roots and accounts for the non-Gaussianity of the quantum system \cite{Chabaud_2022}.

For single bosonic modes, with orthonormal basis $\{\ket{n}\}_{n\in\mathbb{N}_0}$, we can encode our state using the canonical coherent states as,

\begin{equation}
    \ket{z}_{\infty} = \sum_{n\geq 0} \frac{z^n}{\sqrt{n!}} \ket{n}.
\end{equation}

One can treat these as phase-space wave functions of a corresponding quantum state. Instead of representing quantum states as infinite countable vectors as seen in the Fock state description, they can be characterized as holomorphic functions through the transformation,

\begin{equation}\label{transformation}
    \ket{n} \leftrightarrow \qty(z \mapsto \frac{z^n}{\sqrt{n!}}),
\end{equation}

for all $n\in \mathbb{N}_0$. Hence a particular quantum state, decomposed into its Fock basis as $\ket{\psi} = \sum_{n\geq 0} \psi^{(n)} \ket{n}$ can be transformed as,

\begin{equation}
        \ket{\psi} \leftrightarrow F_{\psi}^\star (z) \coloneqq \sum_{n\geq 0} \frac{\psi^{(n)}}{\sqrt{n!}} z^n,
\end{equation}

which is called the stellar function of the state $\ket{\psi}$. This stellar function corresponds to an expansion as a sum in the overcomplete basis of Glauber canonical coherent states \cite{Chabaud_2022}. Using the Hadamard-Weirstrass factorisation theorem, we can rewrite these stellar functions as 

\begin{equation}
    F_{\psi}^\star (z) = e^{-\frac{1}{2} a z^2+b z+c} z^k \prod_n\left(1-\frac{z}{\lambda_n}\right) e^{\frac{z}{\lambda_n}+\frac{1}{2} \frac{z^2}{\lambda_n^2}},
\end{equation}

where the constants, $a,b,c,k,\lambda_n \in \mathbb{C}$ are each dependent on $\ket{\psi}$. Here, $n$ is given as the so-called \emph{stellar rank} of the function, which provides a notion of quantumness as Gaussian states have a stellar rank of zero. For stellar functions of finite rank (i.e. finite roots of the polynomial, $n$), we can write our function as separable in Gaussian and polynomial functions as,

\begin{equation}
    F_{\psi}^\star (z) = G(z)P(z).
\end{equation}

This decomposition can be written as,

\begin{equation}\label{mapping}
    F_{\psi}^\star (z) = e^{-a/2 z^2 + bz+c}\sum_{j=0}^{n} \beta_j z^j,
\end{equation}

for $a,b,c,\beta_j \in \mathbb{C}$, which is the form that we will use in the remainder of the paper.

Stellar functions live in the Segal-Bargmann space, the separable infinite-dimensional Hilbert
space of holomorphic functions $F^\star$ over $\mathbb{C}^m$, satisfying the normalization condition,

\begin{equation}
    \left\|F^{\star}\right\|^2 \ce \frac{1}{\pi^m}\int_{\boldsymbol{z} \in \mathbb{C}^m} \ddd^{2m} z\ \ec^{-\Abs{\bd{z}}} \left|F^{\star}(\boldsymbol{z})\right|^2 <+\infty,
\end{equation}
which constrains $\Re(a)>-1$ in Eq.\ \eqref{mapping}. The SB space has the inner product,

\begin{equation}
\left\langle F_1^{\star} \mid F_2^{\star}\right\rangle_{\tx{SB}} = \frac{1}{\pi^m} \int_{\boldsymbol{z} \in \mathbb{C}^m} \ddd^{2m}z\ \ec^{-\Abs{\bd{z}}} F_1^{\star}(\boldsymbol{z})^* F_2^{\star}(\boldsymbol{z}).
\end{equation}

In the SB space, our operators are functions of the creation and annihilation operators acting on the Hilbert space of our quantum states, and are mapped to differential operators in the SB space by,

\begin{equation}
    \hat{a}^\dagger \leftrightarrow z\times \text{ 
and   } \hat{a} \leftrightarrow \partial_z,
\end{equation}

where $z\times$ acts on a holomorphic function by multiplying it by $z$ and $\partial_z$ takes the partial derivative of it with respect to $z$. It follows that any unitary evolution acting on an element of the SB space remains within the space. %

Common examples of zero stellar rank functions are vacuum states, coherent states, squeezed states and two-mode squeezed states. Fock states of $n$ particle number have stellar rank $n$. Important properties of the stellar rank as a measure of non-Gaussianity include the fact that it is conserved under Gaussian operations, that the states of finite stellar rank form a dense subset of the SB space, and that operationally one can climb the hierarchy by acting on a given state with a creation operator. %
States with infinite rank, such as GKP \cite{Gottesman2001} or cat states, are outside the stellar hierarchy and do not have an obvious measure of quantumness. However, they can be approximated with arbitrary precision using finite-rank states \cite{Chabaud2021}.

We will next use these representations of CV quantum states to develop analytic representations of CV quantum kernels.

\subsection{CV quantum feature maps}\label{CV Quantum Feature Maps section}

Given some metric space of our data $\mathcal{X}$ we can define our CV holomorphic kernel as follows. Firstly, let us encode our data $\txx_i \in \mathcal{X}$ to a pure quantum state which we then decompose into its Fock basis,

\begin{equation}
    \txx_i \to \ket{\psi_{\txx_i}} = \sum_{n\geq 0} \psi_{\txx_{i}}^{(n)} \ket{n}. 
\end{equation}

Using the transformation from equation \ref{transformation} we yield, 

\begin{equation}
        \ket{\psi_{\txx_i}} \leftrightarrow F_{\txx_i}^\star (z) \coloneqq \sum_{n\geq 0} \frac{\psi_{\txx_{i}}^{(n)}}{\sqrt{n!}} z^n.
\end{equation}

This allows our data to be encoded into some continuous variable state via holomorphic function, 
which forms our data encoding,

\begin{equation}
    \txx_i \mapsto \Phi(\txx_i) \coloneqq F_{\txx_i}^\star (z).
\end{equation}

From this, we provide a natural definition of the CV quantum kernel as, 
\begin{equation}
    k(\txx_1,\txx_2) = \Abs{\Braket{\psi(\txx_1) | \psi(\txx_2)}}^2 = \Abs{\Braket{F_{\txx_1}^\star(z) | F_{\txx_2}^\star(z)}_{\tx{SB}}}^2. %
    \label{eq:CVkernel}
\end{equation}
We note that this function is positive semi-definite (appendix \ref{inner products}) and symmetric \cite{schuld_PhysRevLett.122.040504} and thus a valid kernel.

The evaluation of such a kernel can be shown to be equivalent to a CV sampling computation \cite{Chabaud_GcoreComplexity}. Such computations can be exactly simulated in $\mathcal{O}(2^n)$ time provided there are two or more modes \cite{Chabaud_GcoreComplexity,Chabaud_2023}. Using this notion of computational hardness we extend the concept of stellar rank from quantum states to the kernel itself.

Finally, we note that one can define the RKHS as the completion of the span of the kernel function for some data set $\mathcal{X}$. %
We see that the Segal-Bargmann space can be understood as the RKHS of the Gaussian kernel which itself is universal (appendix \ref{Segal Is RKHS}).

\section{General CV kernels}\label{General Kernel Section}

We begin our analysis of CV quantum kernels by considering the general multi-mode case, as any quantum advantage will require $m\ge2$. 

A general $m$-mode state of total stellar finite rank $n$ can be represented by the holomorphic function
\eqn{
	F^{\star}(\bd{z}) = G(\bd{z})P(\bd{z})
        \label{eq:GeneralEncoding0}
}
where $\bd{z}=(z_1,z_2,\dotsb,z_m)^{\trps}$, $G(\bd{z})$ is a Gaussian and $P(\bd{z})$ is a polynomial \cite{Chabaud_2022}.

In general,
\aln{
	G(\bd{z}) &= \exp\lb(-\frac{1}{2} \bd{z}^{\trps} \bd{A} \bd{z} + \bd{B}^{\trps} \bd{z} + C\rb) \nn\\
	P(\bd{z}) &= \sum_{\substack{i_1,i_2,\dotsc,i_m\ge0\\ i_1+i_2+\dotsb+i_m\le n}} \beta_{\bd{i}} z_1^{i_1} z_2^{i_2} \dotsb z_m^{i_m}
    \label{eq:GeneralEncoding}
}

where $\bd{A}\in\bb{C}^{m\times m}$ with components $A_{i,j}$ and $\Re(A_{j,j})>-1$, $\bd{B}\in\bb{C}^m$ with components $B_i$, $C\in\bb{C}$ and $\beta_{\bd{i}}\in\bb{C}$ which are labeled by the vector $\bd{i}\ce(i_1,i_2,\dotsc,i_m)$\footnote{In the case of ${m=2}$ and ${n=2}$, the polynomial will be ${P(z_1,z_2) = \beta_{(0,0)} + \beta_{(1,0)}z_1 + \beta_{(0,1)}z_2 + \beta_{(2,0)}z_1^2 + \beta_{(0,2)}z_2^2 +\beta_{(1,1)}z_1z_2}$.}. The actual values of these components will depend on the particular choice of encoding, and will therefore be functions of the data. 

Explicitly, for some $\txx_1,\txx_2\in\mc{X}$, the feature map is, for $j=1,2$,
\aln{
    \Phi(\txx_j) &= F_{\txx_j}^{\star}(\bd{z}) \nn\\
    &= \exp\lb(-\frac{1}{2}\bd{z}^{\trps}\bd{A}(\txx_j)\bd{z} + \bd{B}(\txx_j)\bd{z} + C(\txx_j)\rb) \sum_{\substack{i_1,i_2,\dotsc,i_m\ge0\\ i_1+i_2+\dotsb+i_m\le n}} \beta_{\bd{i}}(\txx_j) z_1^{i_1} z_2^{i_2} \dotsb z_m^{i_m}
    \label{eq:GeneralEncodingj}
}
and the quantum kernel is
\eqn{
    k(\txx_1,\txx_2) = \Abs{\Bk{F_{\txx_1}^{\star}|F_{\txx_2}^{\star}}_{\tx{SB}}}^2 = \frac{1}{\pi^{2m}} \Abs{\int_{\bd{z}\in\bb{C}^m}\ddd^{2m}z\ \ec^{-\Abs{\bd{z}}^2} F_{\txx_1}(\bd{z})^* F_{\txx_2}(\bd{z})}^2.
    \label{eq:MMkernelstartstart}
}

Calculating this inner product (see appendix \ref{sec:InAB}) requires the evaluation of $2m$ integrals, each of the form \aln{
    I_r(a,b) &\ce \intall \ddd x\ \exp\lb(-ax^2+bx\rb) x^r \nn\\
    &= \frac{\sqrt{\pi}}{a^{(r+1)/2}} \exp\lb(\frac{b^2}{4a}\rb) \sum_{j=0}^{r} \gamma_{r,j} \lb(\frac{b}{\sqrt{a}}\rb)^{j} %
}
where
\eqn{
	\gamma_{r,j} \ce \begin{dcases}
		  \frac{1}{2^r} \frac{r!}{\big((r-j)/2)\big)!j!}, & r\equiv j\ (\tx{mod }2) \\
		0, & \tx{otherwise}
	\end{dcases}
        \label{eq:gammagen}
}
are constants, which depend on the integer values of $r\ge0$ and $0\le j\le r$.

This can be done algorithmically, (see appendix \ref{sec:MMKernel}), and we are able to obtain a closed form expression for the general $m$-mode kernel
\aln{
    \Bk{F_{\txx_1}^{\star}|F_{\txx_2}^{\star}}_{\tx{SB}} &= \exp\lb(C(\txx_1)^*+C(\txx_2) + \sum_{j=1}^{2m} \frac{b_{j-1,j}^2}{4a_{j-1,j}}\rb) \sum_{\substack{i_1,\dotsc,i_m\ge0\\ i_1+\dotsb+i_m\le n}}\ \sum_{\substack{j_1,\dotsc,j_m\ge0\\ j_1+\dotsb+j_m\le n}} \beta_{\bd{i}}(\txx_1)^* \beta_{\bd{j}}(\txx_2) \nn\\ 
    &\qd \times \sum_{\bd{p}=\bd{0}}^{\bd{i}} \sum_{\bd{\qew}=\bd{0}}^{\bd{j}} g(\bd{i},\bd{j},\bd{p},\bd{\qew}) \Vast\{\prod_{\ell=1}^{2m-1} \Vast[\sum_{s_\ell=0}^{r_{\ell-1,\ell}} \frac{\gamma_{r_{\ell-1,\ell},s_\ell}}{a_{\ell-1,\ell}^{(r_{\ell-1,\ell}+s_\ell+1)/2}} \nn\\
    &\qqd \times \sum_{t_{\ell}=0}^{s_{\ell}} \frac{s_{\ell}!}{(s_{\ell}-t_{\ell})!} b_{\ell-1,\ell}^{s_{\ell}-t_{\ell}} \sum_{\substack{u_{\ell,\ell+1},\dotsc,u_{\ell,2m}\ge0\\ u_{\ell,\ell+1}+\dotsb+u_{\ell,2m}=t_\ell}} \lb(\prod_{k=\ell+1}^{2m} \frac{d_{\ell-1,\ell,k}^{u_{\ell,k}}}{u_{\ell,k}!}\rb)\Vast] \nn\\
    &\qqqd  \times \lb[\sum_{s_{2m}=0}^{r_{2m-1,2m}} \frac{\gamma_{r_{2m-1,2m},s_{2m}}}{a_{2m-1,2m}^{(r_{2m-1,2m}+s_{2m}+1)/2}}\ b_{2m-1,2m}^{s_{2m}}\rb]\Vast\}
    \label{eq:MMkernel}
}
where
\aln{
    g(\bd{i},\bd{j},\bd{p},\bd{\qew}) &\ce \prod_{k=1}^{m} {i_k \choose p_k} {j_k \choose \qew_k} (-\iu)^{p_k} (\iu)^{\qew_k}
    \label{eq:gdef}
}
and $a$, $b$, $d$, and $r$ are defined recursively as 
\aln{
	a_{i,j} &\ce a_{i-1,j}-\frac{d_{i-1,i,j}^2}{4a_{i-1,i}} \nn\\
	b_{i,j} &\ce b_{i-1,j}+\frac{b_{i-1,i} d_{i-1,i,j}}{2a_{i-1,i}} \nn\\
	d_{i,j,k} &\ce d_{i-1,j,k} + \frac{d_{i-1,i,j} d_{i-1,i,k}}{2a_{i-1,i}}
    \label{eq:Recusion1}
}
which depend on the initial encoding of $\txx_1$ and $\txx_2$ and
\aln{
    r_{i,k} &\ce r_{i-1,k} + u_{i,k}.
    \label{eq:Recusion2}
}
The initial values of these parameters are defined in appendix \ref{sec:MMKernel} (Eqs.\ \eqref{eq:abdStartOdd}, \eqref{eq:abdStartEven}, and \eqref{eq:seeds}).

While the detail of this kernel function is opaque, we note that any CV encoding of a finite stellar rank will always have a kernel of the form of Eq.\ \eqref{eq:MMkernel}, and can be expressed as a product of a Gaussian and an algebraic function of the parameters of the feature map $A_{i,j}(\txx_k)$, $B_{i} (\txx_k)$, $C(\txx_k)$ and $\beta_{\bd{i}}(\txx_k)$ given in Eq.\ \eqref{eq:GeneralEncodingj}. Specifically, the algebraic function is a solution to the polynomial equation of the form $P_0(\bd{x}) =P_2(\bd{x})f(\bd{x})^2$.  Furthermore, as the modulus squared of the inner product between two physically encoded CV quantum states is always bounded between 0 and 1, this means the kernel is always finite. Mathematically, we can also see this by noting that  Eqs.\ \eqref{eq:GeneralEncoding0} and \eqref{eq:GeneralEncoding} are defined on the SB space, the inner product is bounded. 

Given this structure, we can infer some general properties of the CV kernel. The Gaussian term will cause exponential suppression of kernel values beyond a certain threshold, and while this threshold can be manipulated by bandwidth tuning, this will most likely be at the cost of effective stellar rank. In section \ref{Examples Section}, we will show that this is indeed the case for the specific case of the single mode displaced Fock state kernel; however we stress the details will differ for each specific encoding.

In general, when $m\ge2$, the strong classical simulability of the general multi-mode kernel will scale exponentially on the order of $\mathcal{O}(2^{n})$ \cite{Chabaud_GcoreComplexity,Chabaud_2023}.  We also find that in the case of $\bd{A}\ne\bd{0}$, calculating Eq.\  \eqref{eq:MMkernel} also scales exponentially in $m$, the mode of the encoding, given that $m\leq n$. The scaling with $m$ comes from the general form within the product of sums, 
\begin{equation}
    \prod_{\ell=1}^{2m-1}(\dots),
\end{equation}
representing a deeply nested sum of depth $2m-1$.  Due to the recursion relation of the $r_{\ell-1,\ell}$'s, (Eq.\ \eqref{eq:Recusion2}), each of the sums within this product are dependent on the index of prior sums and the total length of each sum also increases as a function of $m$. It can be easily seen that the computational complexity of $m$ nested sums each of length $\geq \ell$ scales as $\mathcal{O}(\ell^m)$. As such, we see that our general kernel's classical simulability scales as at least $\mathcal{O}(\ell^m)$ for some $\ell > 1$. 

This is a useful property as in practice; an easy way of increasing the stellar rank of a CV quantum feature map is to increase the number of modes rather than directly increasing the stellar rank of a single mode.

Finally, we consider the case of infinite stellar rank. CV states of infinite stellar rank can be approximated arbitrarily well in trace distance\cite{Chabaud_2022}:
\eqn{
    T\big(\Kb{\psi}{\psi},\Kb{F}{F}\big) \le \epsilon
}
where we use $\Ket{\psi}$ to denote a state of infinite stellar rank and $\Ket{F}$ to denote a state of finite stellar rank. Since we only consider pure states, the trace distance can be easily expressed in terms of the inner product:
\eqn{
    T\big(\Kb{\psi}{\psi},\Kb{F}{F}\big) = \sqrt{1-\Abs{\Bk{\psi|F}}^2} \le \epsilon \imp 1-\Abs{\Bk{\psi|F}}^2 \le \epsilon^2.
    \label{TDClosesess}
}
From this fact, it can be shown (details in appendix \ref{sec:infinitekernels}) that given two states of infinite stellar rank $\Ket{\psi_1}$ and $\Ket{\psi_2}$, and two states of finite stellar rank $\Ket{F_1}$ and $\Ket{F_2}$ such that
\aln{
    \Abs{\Bk{\psi_1|F_1}}^2 &\ge 1-\epsilon^2 \nn\\
    \Abs{\Bk{\psi_2|F_2}}^2 &\ge 1-\epsilon^2
    \label{eq:IPCloseness}
}
for some small $\epsilon$, then
\eqn{
    \Big|\Abs{\Bk{\psi_1|\psi_2}}^2 - \Abs{\Bk{F_1|F_2}}^2\Big| \le 4\sqrt{2}\epsilon.
    \label{eq:KernelCloseness}
}
In other words, kernels defined by CV feature maps of infinite stellar rank can be approximated arbitrarily well by kernels of the form of Eq.\ \eqref{eq:MMkernel}, which is defined by feature maps of finite stellar rank.

\section{Displaced Fock state kernel}\label{Examples Section}

In this section, we proved a simple example of a CV quantum kernel.
We construct an analytic expression for a kernel generated from a displaced, single-mode
Fock state encoding which we show this is rotationally and translationally invariant, radial and characteristic. 
While for a single mode encoding, we do not have exponential growth in simulablity with stellar rank, we will use stellar rank to generate intuition for the multi-mode encoding.

\subsection{Closed form \& analytic properties}

We first consider a simple single-mode bosonic kernel which encodes data through a general Fock state $\ket{n}$ with an applied displacement unitary $\hat{D}(\alpha) \ket{n}$. To encode 2 pieces of information $\boldsymbol{\alpha}=(\alpha_1,\alpha_2)^\trps \in \mc{X} \subset \bb{R}^2$, we parameterise the displacement operator by the complex number $\alpha\ce\alpha_1 + i \alpha_2$. This operator acts on any holomorphic function as 
\begin{equation}
    F^{\star}(z) \mapsto \mathrm{e}^{\alpha z-|\alpha|^{2} / 2} F^{\star}\left(z-\alpha^{*}\right).
\end{equation}
Since displacement is a Gaussian operation, it will not change the stellar rank of $F^{\star}(z)$; therefore the encoded state $\hat{D}(\alpha) \Ket{n}$ will have a stellar rank of $n$. Explicitly, the encoded state is 

\begin{equation}
	\hat{D}(\alpha)\Ket{n} \leftrightarrow F_{\alpha}^{\star}(z) = \ec^{\alpha z - \Abs{\alpha}^2/2} \frac{(z-\alpha^*)^n}{\sqrt{n!}}
 \label{eq:Dencode}
\end{equation}
and with this encoding, the quantum kernel is
\eqn{
	k(\bd{\alpha},\bd{\beta}) = \Abs{\Bk{F_{\alpha}^{\star}(z)|F_{\beta}^{\star}(z)}_{\tx{SB}}}^2
}
where
\aln{
    \Bk{F_{\alpha}^{\star}(z)|F_{\beta}^{\star}(z)}_{\tx{SB}} &= \int_{z\in\bb{C}} \ddd^2 z\ \ec^{-\Abs{z}} \big(F_{\alpha}^{\star}(z)\big)^* F_{\beta}^{\star}(z) \nn\\ 
    &= \frac{1}{\pi} \frac{\ec^{-(\Abs{\alpha}^2+\Abs{\beta}^2)/2}}{n!} \int_{z\in\bb{C}} \dd^2z\ \ec^{-(\Abs{z}^2-\alpha^*z^*-\beta z)} (z^*-\alpha)^n (z-\beta^*)^n.
    \label{eq:Dkernelstart}
}

After the integration and some algebra (see appendix \ref{sec:KDdetails} for details), we can write down the displacement kernel in closed form as:
\aln{
    k_{D}(\bd{\alpha},\bd{\beta}) &= \Abs{\Bk{F_{\alpha}^{\star}(z)|F_{\beta}^{\star}(z)}_{\tx{SB}}}^2 \nn\\
    &= (n!)^2 \ec^{-\Abs{\alpha-\beta}^2} \Bigg|\sum_{i=0}^{n} \sum_{j=0}^{n-i} \sum_{k=0}^{n} \sum_{\ell=0}^{n-k} \sum_{p=0}^{i+k} \sum_{\qew=0}^{j+\ell} \frac{(-\iu)^j (-\alpha)^{n-i-j}}{i! j! (n-i-j)!} \frac{(\iu)^{\ell} (-\beta^*)^{n-k-\ell}}{k! \ell! (n-k-\ell)!} \nn\\
    &\qqd \times \gamma_{(i+k),p} \gamma_{(j+\ell),\qew} \big(\alpha^*+\beta\big)^p \big(-\iu(\alpha^*-\beta)\big)^\qew\Bigg|^2
    \label{eq:Dkernel}
}
where the $\gamma_{r,j}$'s are defined in Eq.\ \eqref{eq:gammagen}.
In appendix \ref{sec:KDExamples}, we list the explicit form of this kernel for the first 9 Fock states, which can be used to directly calculate the kernel entries via the kernel trick. However beyond $n=9$, and for more complex kernels, classical simulation quickly becomes intractable.

From Eq.\ \eqref{eq:Dkernel}, it is clear that the displacement kernel is the product of a Gaussian and a polynomial of degree $4n$ in both $\alpha$ and $\beta$, because
\alns{
    &2\big(i+k+j+\ell+(n-i-j)\big) = 2\big(n+k+\ell\big) \nn\\
    &\qqd \xrightarrow[\ell]{\max} 2\big(n+k+(n-k)\big) = 4n \nn\\
    &2\big(i+k+j+\ell+(n-k-\ell)\big) = 2\big(n+i+j\big) \nn\\
    &\qqd \xrightarrow[j]{\max} 2\big(n+i+(n-i)\big) = 4n
}

Using this closed form expression we are also able to show that the kernel is translation (shift) and rotation invariant (see appendices \ref{sec:KDShift} and \ref{sec:KDRotation}). From these properties, we find that (see appendix \ref{sec:KDRadial})
\eqn{
    k(\bd{\alpha},\bd{\beta}) = k(\Abs{\bd{\alpha}-\bd{\beta}})
    \label{eq:KDradil},
}
and so it is a radial kernel. This combined with the fact that the Fourier transform of this kernel is also the product of a Gaussian and a polynomial of degree $4n$, which has support over the entire Fourier domain, except at a finite number of points (see appendix \ref{sec:KDFT}), means that the displaced Fock state kernel is a characteristic kernel \cite{Sriperumbudur2008}.

\begin{figure}
    \centering
    \includegraphics[width=0.48\textwidth]{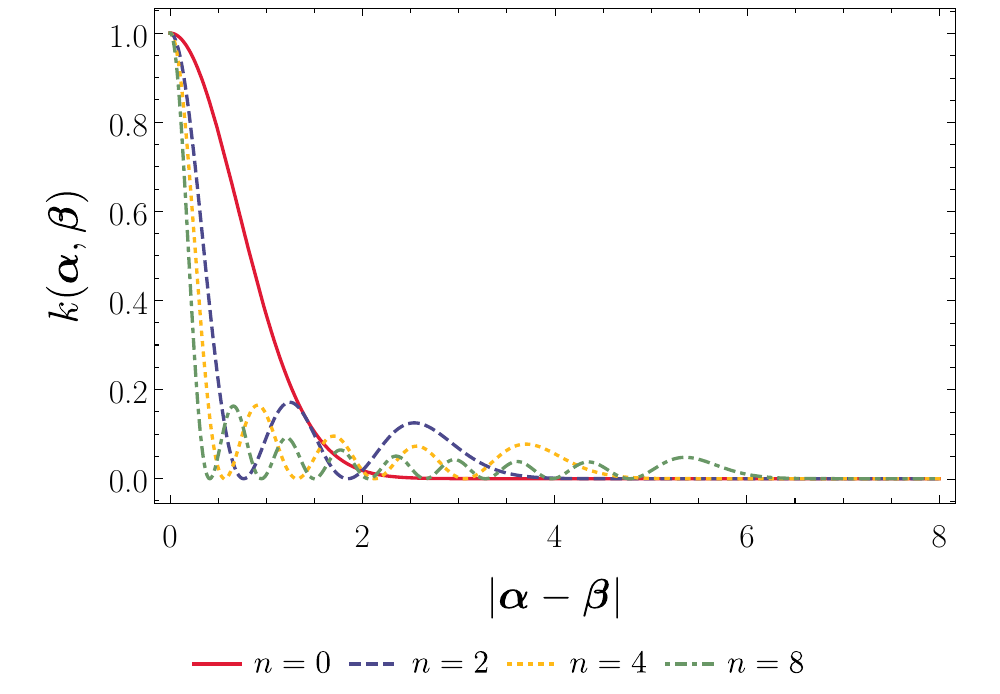}
    \caption{%
            The kernel function in the case of displacement encoding (Eq.\ \eqref{eq:Dencode}) as a function of $\Abs{\bd{\alpha}-\bd{\beta}}$, the distance in the data space, for various values of the initial Fock state, $n$.
        }
    \label{fig:Dkernel}  
\end{figure}

In order to generate some intuition as to how such a kernel will behave, in figure \ref{fig:Dkernel} we plot the displaced Fock state kernel function, $k(\Abs{\bd{\alpha}-\bd{\beta}})$, for training data chosen from a uniform distribution of $\Abs{\bd{\alpha}-\bd{\beta}}$ from 0 to 8, for various values of the stellar rank, $n$. We see in the figure that as the value of $n$ increases, so does the number of zeros in the kernel function, as expected. Additionally, as the stellar rank increases, the kernel's ability to distinguish between large distances in the original data space, $\Abs{\bd{\alpha}-\bd{\beta}}$, improves. For example, the $n=2$ kernel function will evaluate as zero for any distance $\Abs{\bd{\alpha}-\bd{\beta}}$ greater than 4, whereas the $n=8$ kernel will have non-zero values up to a distance of 6. 

We can also see that for kernels of finite stellar rank, due to multiplication by the Gaussian factor $\ec^{-\Abs{\bd{\alpha}-\bd{\beta}}^2}$, there will be a threshold value beyond which all kernel evaluations will be exponentially close to zero. Additionally, it appears that as stellar rank increases, the amplitudes of the maxima diminish. %
This would suggest that as the stellar rank increases, kernel values outside the central peak will become increasingly suppressed.
This is further supported by the fact that the displaced Fock state kernels integrate to a constant value, $\pi$ (appendix \ref{sec:KDintegral}), which limits how large each maximum can be.
Given these two results, we would expect, therefore, that displaced Fock state CV kernels of high stellar rank will generate kernel values that are increasingly concentrated at low values, a feature that we observe in the learning experiments conducted below. Given kernel values are always statistically approximated by repeated measurement, as kernel values become smaller they require more measurements to remain distinguishable. Consequently, we expect high stellar rank models with low kernel variance will also become increasingly vulnerable to shot noise.

\FB

\subsection{Bandwidth tuning}

This observation has some of the flavour of exponential concentration and recent work has proposed bandwidth tuning as a possible mitigation technique \cite{canatar2023bandwidth}. Let us therefore consider the consequences of implementing a bandwidth $c$ on the encoded data by taking
\begin{equation}
    \bd{\alpha} \to c\bd{\alpha},
\end{equation}
where $c>0$ is a real hyperparameter. Physically, this corresponds to a reduction in the value of the displacement of the Fock state, $\hat{D}(\alpha)\Ket{n} \to \hat{D}(c\alpha)\Ket{n}$, which results in a non-linear transformation of the kernel. 

\begin{figure*}
    \includegraphics[width=0.24\textwidth]{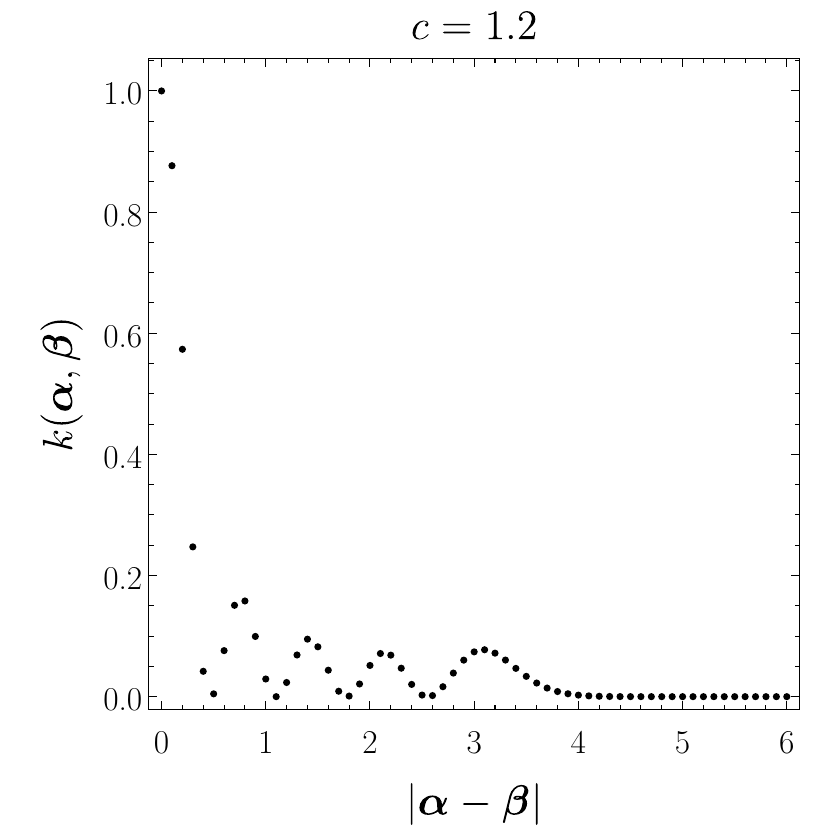}
    \includegraphics[width=0.24\textwidth]{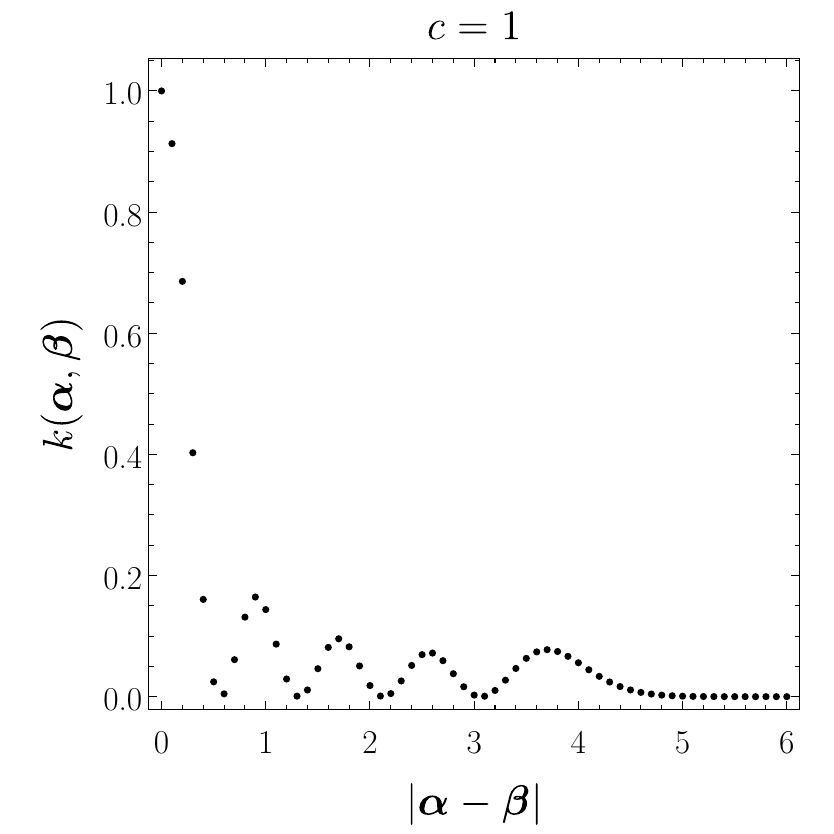}
    \includegraphics[width=0.24\textwidth]{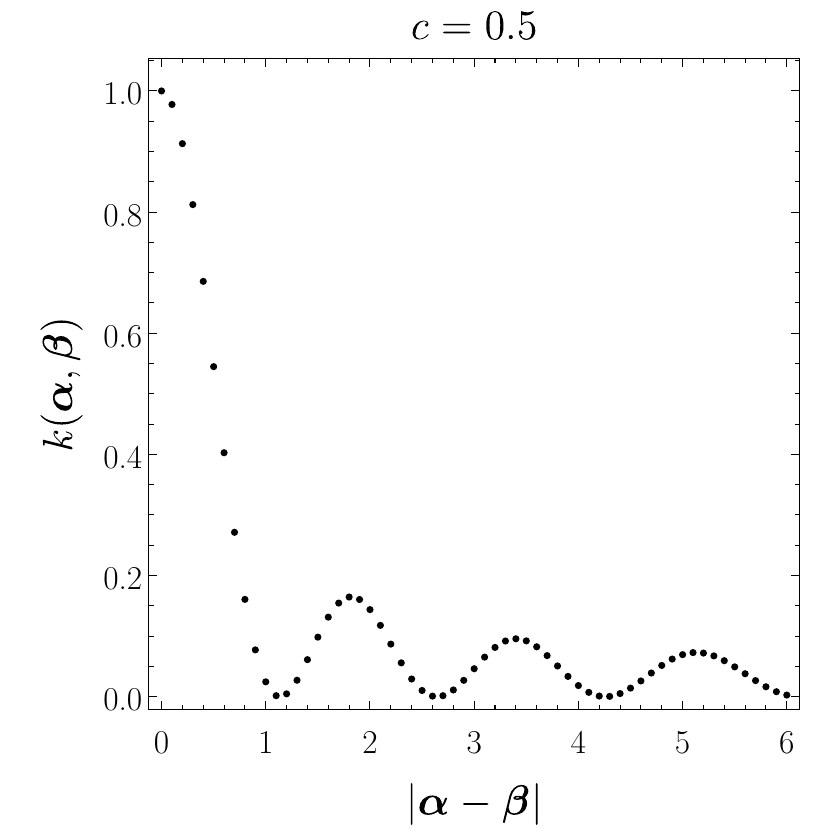}
    \includegraphics[width=0.24\textwidth]{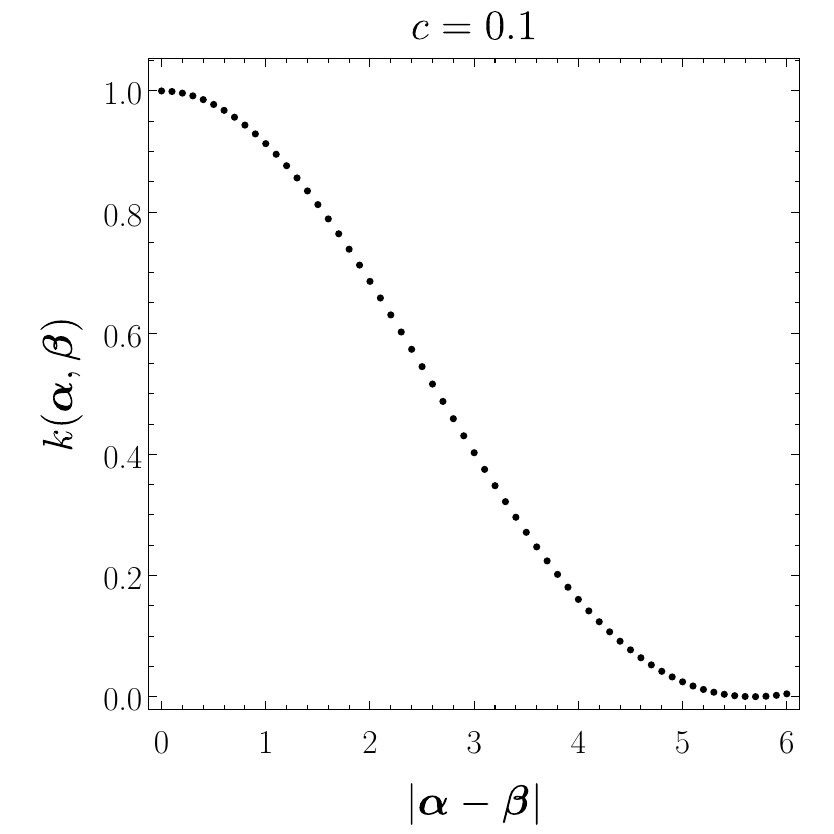}
    \caption{%
            The kernel function in the case of displacement encoding (Eq.\ \eqref{eq:Dencode}) with $n=4$ for a uniform distribution of ${\Abs{\bd{\alpha}-\bd{\beta}}\in [0,6]}$ for various values of the bandwidth hyperparameter set to, from left to right, $c=1.2$, $c=1$, $c=0.5$, $c=0.1$.
        }
    \label{fig:DkernelBandwidth}  
\end{figure*}

In figure \ref{fig:DkernelBandwidth}, we explore the effect of the bandwidth $c$ on the value of the displaced Fock state kernel function. We take some data, which we choose to be a uniform distribution of $\Abs{\bd{\alpha}-\bd{\beta}}$ from 0 to 6, and apply a bandwidth so that
\eqn{
    \Abs{\bd{\alpha}-\bd{\beta}} \to \Abs{c\bd{\alpha}-c\bd{\beta}} = c\Abs{\bd{\alpha}-\bd{\beta}}.
}
We find that for this particular choice of distribution, a bandwidth of $c=0.5$ moves all of the data away from the tail of the function (where values are exponentially close to zero). For example, all data separated in the original space by values $>5$ but $<6$ have kernel values of near zero for $c=1$ (left figure) but for a bandwidth of $c=0.5$, we can see these data points can now be discriminated. Unfortunately, it is also the case that data points that were easily distinguishable for $c=1$, such as the two points closest to the y axis, are less distinguishable following bandwidth tuning to $c=0.5$. Furthermore, we also see that this bandwidth reduces the \emph{effective} stellar rank of the kernel function, since there are now only $n=4$ maxima, rather than $n=5$ over the range of the distribution (although we note that the actual stellar rank of the kernel function does not change, since it is still defined for all $\Abs{\bd{\alpha}-\bd{\beta}}\in[0,\infi)$). On the other hand, a bandwidth is $1.2$, moves more of the data into the tail of the function as compared with no bandwidth, exponentially suppressing data with values larger than $\sim4$. In the case of a very small bandwidth, $c=0.1$, the kernel function is effectively a Gaussian. %
In conclusion, it is reasonable to expect that the hyperparameter, $c$ will need to be carefully chosen for each problem. 

\FB

\subsection{Learning experiments}

\begin{figure}[!htbp]
    \centering
    \includegraphics[width = \textwidth]{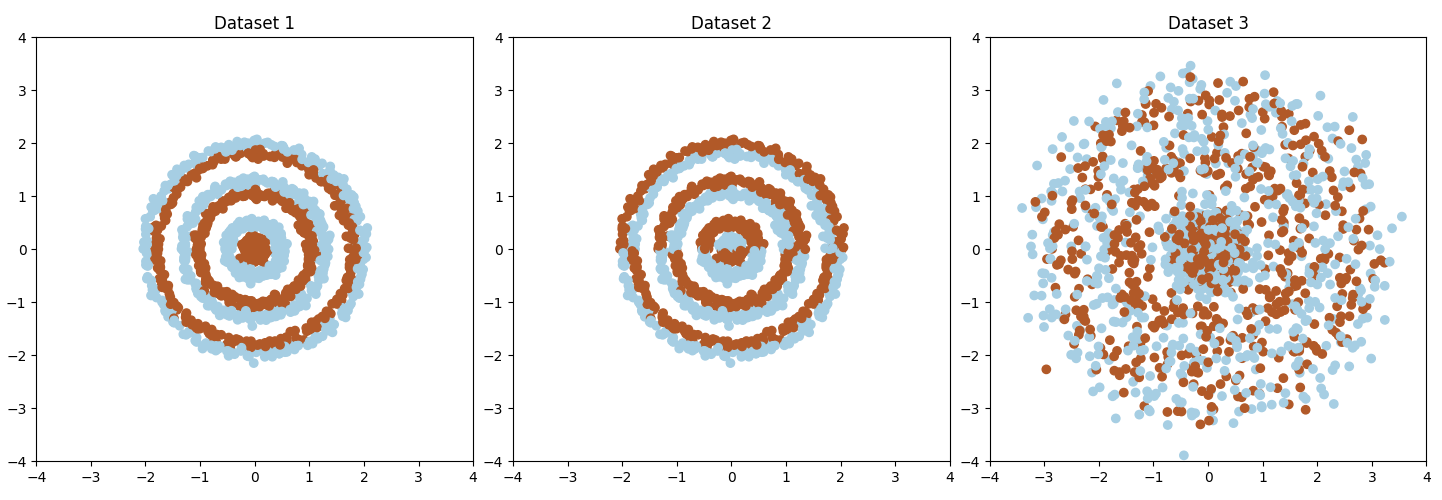}
    \caption{Data set 1 (\textit{left}) includes tightly packed circles with little added noise while data set 3 (\textit{right}) has circles with larger separation and substantially more noise. Data set 2 (\textit{centre}) is a modification of data set 1 where labels above $y=0$ are flipped, which adds more complexity to the data.}
    \label{fig:data shown visually}
\end{figure}

In this section we present some performance metrics and decision boundaries for implementations of a range of displaced Fock state kernels. Inspired by previous work showing the advantage of matching the structure of kernel with that of the data \cite{glick2022covariant,Liu_2021, ragone2023representation}, we create a task to exploit the underlying structure (radial symmetry) of the displaced Fock state kernel\footnote{Our source code can be found at \href{https://github.com/rishigoel2003/Quantum-Kernel-Machine-Learning-with-Continuous-Variables}{https://github.com/rishigoel2003/Quantum-Kernel-Machine-Learning-with-Continuous-Variables}}. The task is a supervised learning classification task using an annular data set constructed by combining multiple instances of the Scikit-learn data set method, \textit{make circles}. We combine 3 instances of the method, each with different parameters to yield 3 concentric circles of data with binary labels. We define 500 data points in the data set for each set of circles, and an equal number for each label $\{0,1\}$. We create three data sets: one with circles that are close together and a small amount of noise (0.05), one which modifies the first by flipping the labels above the $y=0$ axis, and one which modifies the first by increasing the separation between the circles and adding more noise (0.3). The three data sets are illustrated in figure \ref{fig:data shown visually}. When we increase the separation between the circles in the third data set, we maintain the ratios of radii of the blue and red circles as 0.3 for the inner circles, 0.8 for the middle circles and 0.9 for the outer circles. In all three learning simulations, the train-test split is the default 75\%, 25\% split.

\begin{figure}[!htbp]
    \centering
    \includegraphics[width=\textwidth]{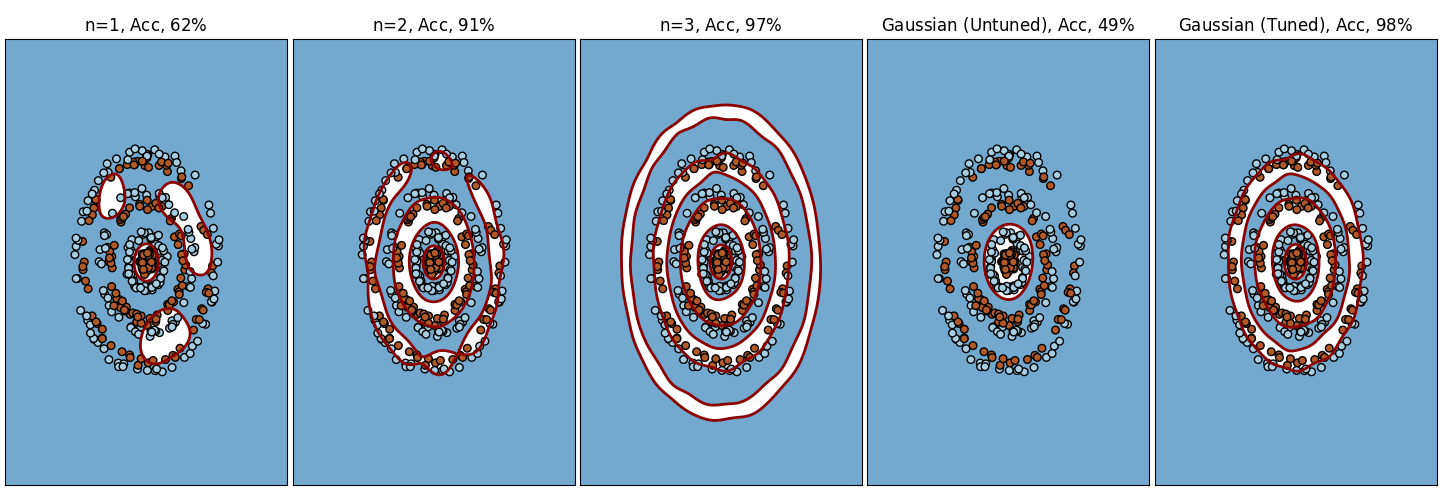}
    \caption{Plots of the decision boundaries for $n=1,2,3$ displaced Fock state kernels (left three panels) with no hyperparameter tuning, benchmarked against a classical Gaussian kernel (right two panels) for data set 1. As this annular data is constructed from close circles, we see that accuracy increases with stellar rank as higher rank kernels can identify finer structure. On this particular problem, the quantum displaced Fock state kernels outperform the Scikit-learn Gaussian kernel with default hyperparmeters. When the Gaussian kernel is tuned using Bayesian optimisation, it can fit the data to a much higher accuracy than when it is untuned.}
    \label{fig:displacement kernel comparisons}
\end{figure}

\FB 
First, we test on data set 1.
In figure \ref{fig:displacement kernel comparisons} we plot the test data and decision boundaries and accuracy for five kernels, three displaced Fock state kernels with $n=1,2,3$, the Scikit-learn Gaussian kernel without hyperparmeter tuning and the same Gaussian kernel which has been tuned using Bayesian optimization. We do not perform any tuning on the displaced Fock state kernels. We see that the accuracy of the displaced Fock state kernels improves with increasing stellar rank (first three panels) and the $n=2$ and $n=3$ kernels significantly outperform the default classical Gaussian kernel (fourth panel). However, when the Gaussian kernel has been tuned, it is also able to classify the data to a high degree of accuracy (fifth panel). Tuning the Gaussian kernel via Bayesian optimisation to achieve this accuracy takes a significant amount of computational time, while the displaced Fock state kernels require no hyperparameter tuning, suggesting that these quantum kernels are better suited for this task.

\FB

\begin{figure}
    \centering
    \includegraphics[width=\textwidth]{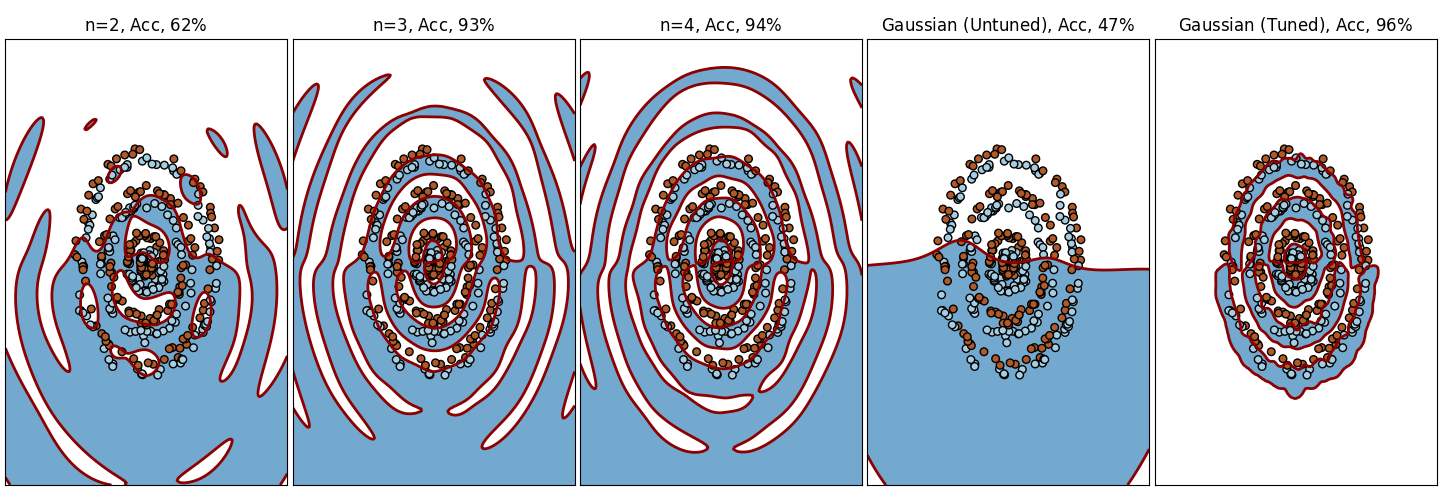}
    \caption{Plots of the decision boundaries for $n=2,3,4$ displaced Fock state kernels (left three panels) with no hyperparameter tuning, benchmarked against a classical Gaussian kernel (right two panels) for data set 2. The accuracy increases with stellar rank. The quantum displaced Fock state kernels outperform the Scikit-learn Gaussian kernel with default hyperparmeters; however, after the Gaussian kernel is tuned using Bayesian optimisation, it can fit the data to a much higher accuracy than when it is untuned.}
    \label{fig:crazy circles dataset}
\end{figure}

In data set 2, we add an additional complexity by flipping the labels in data set 1 above the $y=0$ axis, and again test on five kernels: the displaced Fock state kernel with stellar rank $n=2,3,4$ with no hyperparameter tuning, and the Gaussian kernel without and with hyperparameter tuning. The decision boundaries, test data and accuracy of these kernels is  plotted in figure \ref{fig:crazy circles dataset}. Similar to the tests on data set 1, we find that the accuracy of the displaced Fock state kernels improves with increasing stellar rank (first three panels), since higher stellar rank kernels can identify finer structure. In particular, the $n=3$ and $n=4$ kernels were well suited for this learning task. The $n=1$ kernel had very poor accuracy, so it was omitted from this figure. We also found that the Gaussian kernel also performed well, but again only after tuning via computationally expensive Bayesian optimisation (fourth and fifth panel), suggesting it is less suited for such a classification task.

\begin{figure}[!htbp]
    \centering
    \includegraphics[width=\textwidth]{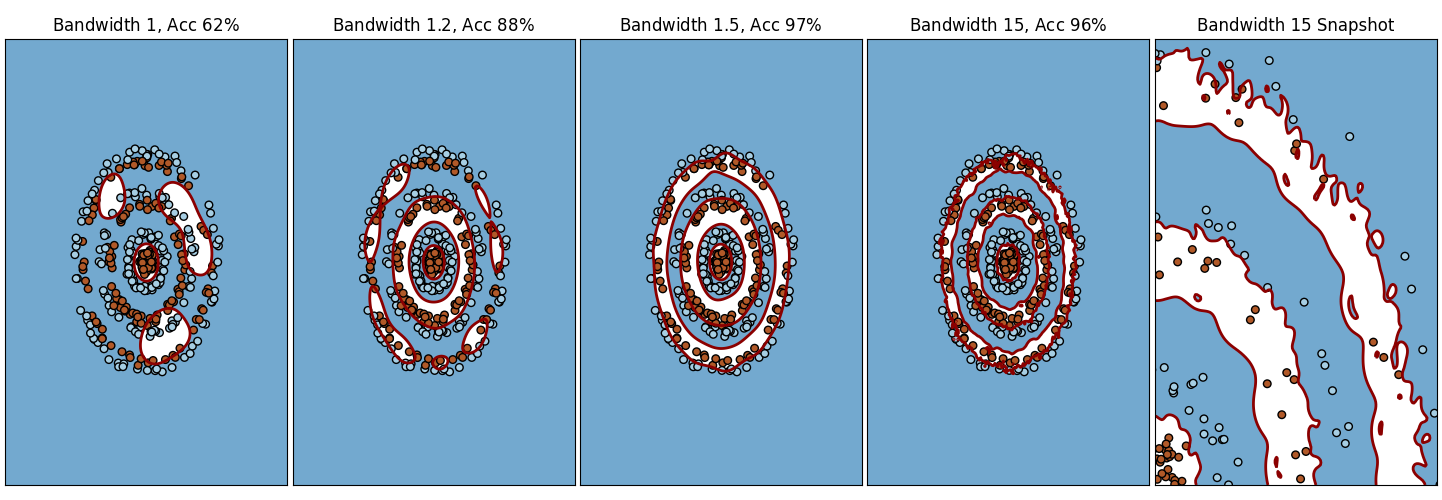}
    \caption{Here we use data set 1 and plot the performance of the $n=1$ displaced Fock state kernel. Tuning the bandwidth hyperparameter permits the kernel to learn the finer grained detail of the data and will improve the performance of an underfit model. However, further tuning of bandwidth can eventually result in a model that is overfit (right-most panel).}
    \label{fig:increase bandwidth displacement}
\end{figure}

Next, we consider the effect of the bandwidth hyperparmeter on the displaced Fock state kernels. In figure \ref{fig:increase bandwidth displacement}, we use data set 1 to determine the effect of the bandwidth on the accuracy of the $n=1$ displaced Fock state kernel, which under-fits the data.
Increasing the bandwidth from 1 to 1.5 results in an improvement in accuracy from 62\% to 97\%. However, increasing bandwidth to 15 results in over-fitting and test accuracy starts to decline again.

\begin{figure}[!htbp]
    \centering
    \includegraphics[width=\textwidth]{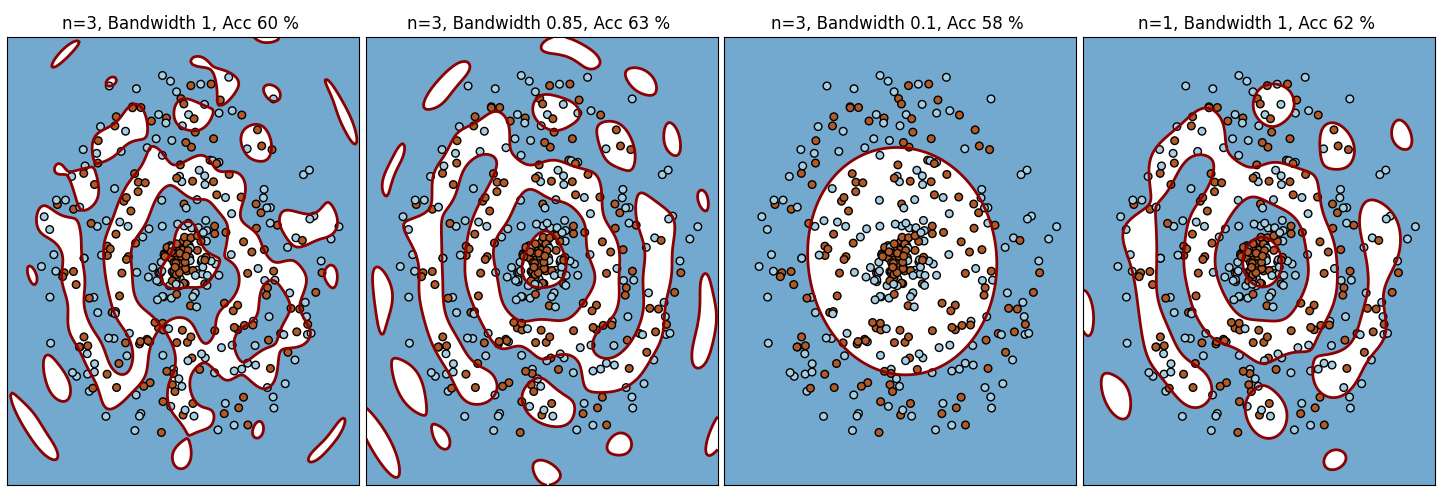}
    \caption{In the case where data is noisy, with wider and further separated circles, we test the $n=3$ kernel against 3 values of bandwidth. We see that reducing the bandwidth can take an over-fit kernel (panel 1) to one which generalises better (panel 2). The tradeoff, however, is that the decision boundary now closely resembles one for the lower rank $n=1$ kernel (panel 4). Again, further tuning leads to a Gaussian kernel (panel 3).}
    \label{fig:decrease bandwidth displacement}
\end{figure}

In figure \ref{fig:decrease bandwidth displacement} we use data set 3 to determine how the bandwidth affects generalisation for noisy data. We see that for the $n=3$ while bandwidth tuning can improve the performance from 60\% to 63\%, the decision boundary becomes closer to that of a lower rank kernel (last panel). Continued bandwidth tuning eventually results in a kernel that approximates an un-tuned Gaussian kernel. This corroborates the theoretical behaviour we illustrated in figure \ref{fig:DkernelBandwidth}: decreasing the bandwidth decreases the effective stellar rank of the kernel.

Overall, these learning simulations demonstrate the displaced Fock state CV kernel is well suited to learn on data sets whose structure matches the symmetry of kernel function (e.g. annular data). Additionally, tuning the bandwidth hyperparemeter can improve accuracy if the model is under-fit, or the data is noisy.

\FB

\section{Qudit kernels}
\label{Qudit Kernel Section}

Thus far we have only examined the CV case and it is interesting to ask if any of our results are applicable in the more familiar discrete qubit/qudit case. In fact, qudits of dimension $d$ can also be represented as complex polynomials in the SB space as \cite{Chabaud_2022}
\eqn{
    \Ket{\psi} = \sum_{j=0}^{d-1} \alpha_j \Ket{j} \to F^{\star}_{\psi}(z) = \sum_{j=0}^{d-1} \ms{n}_{d,j} \alpha_j z^j
    \label{eq:Quditstate}
}
where $\ms{n}_{d,j}\in\bb{C}$ is a normalization factor, which only depends on $d$ and $j$ and ensures that
\eqn{
    \Bk{F^{\star}_{\psi}(z)|F^{\star}_{\psi}(z)}_{\tx{SB}} = \sum_{j=0}^{d-1} |\alpha_j|^2 = 1.
    \label{eq:QuditNormailzation}
}
In appendix \ref{sec:QuditNormalization}, we show Eq.\ \eqref{eq:QuditNormailzation} requites that
\eqn{
    \ms{n}_{d,j} = \frac{1}{\sqrt{j!}}
}
up to a global phase.

We note that a tensor product of $m$ qudits, each of dimension $d$, can always be written as a single qudit of dimension $m\times d$, so we will only consider the single-mode case.

As with the CV kernels, we consider the qudit kernel to be
\eqn{
    k(\txx_1,\txx_2) = \Abs{\Bk{\psi(\txx_1)|\psi(\txx_2)}}^2
}
where the data $\txx_1, \txx_2 \in \mc{X}$ are encoded into the the states $\Ket{\psi(\txx_1)}$ and $\Ket{\psi(\txx_2)}$ respectively. In the case of mixed states, where the data is encoded into a density matrix $\hat{\rho}(\txx) \in \bb{C}^d\otimes\bb{C}^d$, we will consider the vectorisation, which stacks the columns of the matrix to form as single vector $\Ket{\rho(\txx)}\rangle \in \bb{C}^{2d}$ \cite{schuld2021-quantumkernel}. The vectorised state can then be represented as Eq.\ \eqref{eq:Quditstate}.

With the encoded qudits represented as polynomials in the SB space, we can calculate the inner product as (see appendix \ref{sec:Quditcalc}):
\aln{
    \Bk{\psi(\txx_1)|\psi(\txx_2)} &\to \Bk{F^{\star}_{\txx_1}(z)|F^{\star}_{\txx_2}(z)}_{\tx{SB}} \nn\\
    &\quad = \frac{1}{\pi} \int_{z\in\bb{C}} \ddd^2 z\ \ec^{-z^2} \vast(\sum_{i=0}^{d-1} \frac{\alpha_i(\txx_1)^*}{\sqrt{i!}} (z^*)^i\vast)\lb(\sum_{j=0}^{d-1} \frac{\alpha_j(\txx_2)}{\sqrt{j!}} z^j\rb) \nn\\
    &\quad = \sum_{j=0}^{d-1} \alpha_j(\txx_1)^* \alpha_j(\txx_2).
    \label{eq:QuditKernel}
}
as one would expect.

In appendix \ref{sec:Quditcalc2}, we show this same expression can also be calculated from the general multi-mode kernel (Eq.\ \eqref{eq:MMkernel}, by setting $m=1$, $n=d-1$ and $\beta_j=\alpha_j/\sqrt{j!}$, showing that \emph{all} quantum kernels that can be written as Eq.\ \eqref{eq:GeneralQuantumKernel} can be written as either an algebraic function, a Gaussian, or the product of a Gaussian and an algebraic function.

\section{Conclusions \& future work}\label{Conclusions and future work section}

In this paper we use the holomorphic representation of continuous variable quantum states to present a closed form kernel which describes how one might encode data for CV quantum kernel machine learning. In doing so, we are able to identify that all quantum kernels of finite stellar rank can be expressed analytically as products of Gaussian and algebraic function of the parameters of the feature map, $A_{i,j}(\txx_k)$, $B_{i} (\txx_k)$, $C(\txx_k)$ and $\beta_{\bd{i}}(\txx_k)$. In the case of two or more modes,
the measure of stellar rank, a quantity that is easy to characterise for practical bosonic implementations, neatly captures the classical hardness of simulating these kernels, which scales as $\mathcal{O}(2^{n})$. We then prove kernels defined by feature maps of infinite stellar rank, such as GKP-state encodings, can be approximated arbitrarily well by kernels defined by feature maps of finite stellar rank. Furthermore, by analysing a simple single mode CV kernel which are expressed as the product of a Gaussian and a polynomial, we are able to develop intuition for how multi-mode kernels will behave as we increase their ``quantumness" as measured by stellar rank. We see that it is likely that one will encounter problems analogous to exponential concentration, and while bandwidth tuning may mitigate this to some extent, this will trade-off with maintaining effective stellar rank and  robustness to shot noise. We have also shown that it is possible to construct characteristic CV quantum kernels that embody both translational and rotational invariance, with an explicit example given as a kernel constructed by a displaced Fock state. We leverage this structure, by creating a supervised learning classification task on annular data. We find that displaced Fock state kernel is well suited for this task, and the classification accuracy increases with increasing stellar rank.
While we have not simulated any multi-mode kernels here, we would expect similar behaviour in terms of kernel value suppression beyond a threshold value due to the Gaussian term, and also increasing stellar rank. 

It will be important in future work to quantitatively characterise the bandwidth generated trade-offs for particular target problems and specific physical implementations. We also leave for future work an analysis to consider the effects of various appropriate noise models. Of relevance is recent work employing Kerr kernels derived from encoding data into the phase and amplitude of Kerr coherent states to construct Kerr quantum kernels. Empirical analysis demonstrates the robustness of Kerr coherent states to various noise models and the authors attribute this to their flexibility in accommodating different hyperparameters \cite{Enhancing_2024}. Finally, we note a concurrent work has recently appeared that uses phase space negativity to estimate the efficiency of the classical simulation of quantum kernel functions for bosonic systems \cite{Chabaud_2024}. It will be interesting in future to understand the connection between this work and the results presented here.

\section*{Acknowledgements}
This work has been supported by the Australian Research Council (ARC) by Centre of Excellence for Engineered Quantum Systems (EQUS, CE170100009). We wish to thank Aleesha Isaacs, Carolyn Wood, Riddhi Gupta, Gerard Milburn, Andrew White and Nicolas Menicucci for useful discussions.

\bibliographystyle{quantum}
\bibliography{citations}%

\clearpage

\begin{center}
{\Large Supplementary Material}
\end{center}

\appendix

\section{Proof that inner products are positive semi-definite}\label{inner products}

\begin{remark}\label{inner product positive}
    Inner Products of the form $\braket{\cdot|\cdot}: \mathcal{H}\times\mathcal{H}\to \mathbb{C}$ are positive semi-definite. 
\end{remark}

    A matrix $\boldsymbol{M}$ is positive semi-definite if and only if 

    \begin{equation}
        \boldsymbol{x}^\dagger \boldsymbol{M} \boldsymbol{x} \geq 0 \qquad \forall \boldsymbol{x} \in \mathbb{C}^n.
    \end{equation}

    We can construct our Gram Matrix, $\boldsymbol{G}$, from our defined inner product by 

    \begin{equation}
        G_{i,j} = \braket{\boldsymbol{v_i}|\boldsymbol{v_j}},
    \end{equation}

    for $\boldsymbol{v_i},\boldsymbol{v_j} \in \mathcal{H}$.

    Hence to show the inner product is positive semi-definite, it is sufficient to show the gram matrix is positive semi-definite. We can see for any $\boldsymbol{x} \in \mathbb{C}^n$,

    \begin{align}
        \boldsymbol{x}^\dagger \boldsymbol{G} \boldsymbol{x} & = \sum_{i,j} x_i^* G_{i,j} x_j\\
        &= \sum_{i,j} x_i^* \braket{\boldsymbol{v_i}|\boldsymbol{v_j}} x_j \\
        &= \sum_{i,j}\braket{x_i^*\boldsymbol{v_i}|x_j\boldsymbol{v_j}} \\
        &= \Bk{\sum_i x_i^*\boldsymbol{v_i}| \sum_j x_j\boldsymbol{v_j}}\\
        &= \norm{\sum_i x_i^*\boldsymbol{v_i}}^2 \geq 0 
    \end{align}

\begin{remark}\label{conjugate symmetry}
    Note that by definition, inner products are also conjugate symmetric. That is, 

    \begin{equation}
        \braket{x|y} = \braket{y|x}^*.
    \end{equation}
\end{remark}

\section{Segal-Bargmann space is a RKHS of the Gaussian kernel}
\label{Segal Is RKHS}

Our formulation of CV quantum kernels heavily utilises the Segal-Bargmann space. Knowing properties about this space can allow us to further develop our kernels. Here we explicitly prove that this space contains the Gaussian kernel which is well known to be universal. We can also show that this kernel is universal. Universality of kernels is a measure of their expressibility, characterising how well the kernel can classify the data from a metric space. A universal kernel is one which can learn any function for empirical loss minimisation from equation \ref{minimised supervised learning problem}. To prove such a property we can show that the RKHS is dense inside the space of continuous functions ($\mathcal{C}(\mathcal{X})$) of our data. This is equivalent to showing that the kernel, decomposed into a power series of inner products, has only positive coefficients \cite{lecture_notes}. 

The following proof is adapted from \cite{Bargmann_1961}. Let us denote our reproducing kernel as $\textbf{e}_a \in \mathcal{H}_{SB}$ for some fixed $a\in \mathcal{C}^n$. Hence by the reproducing property (Eq.\eqref{eq:repprop}) we have,

\begin{equation}\label{reprod}
    f(a) = \braket{\textbf{e}_a|f}.
\end{equation}

By the definition of the inner product in the SB space we can rewrite this as,

\begin{equation}
    f(a) = \int_{\mathcal{C}^m} \ddd^{2m} z\ {\textbf{e}^*_{a}(\boldsymbol{z})} f(\boldsymbol{z}) .
\end{equation}

Here we can see that the reproducing kernel ${\textbf{e}^*_{a}(\boldsymbol{z})} = \mathcal{K}(a,z)$ is the kernel function for the SB space. By definition (Eq.\ \eqref{reprod}) we have, 

\begin{equation}
    \textbf{e}_a(z) = \braket{\textbf{e}_z|\textbf{e}_a},
\end{equation}

and so we yield, 

\begin{equation}
    \mathcal{K}(a,z) = \braket{\textbf{e}_a|\textbf{e}_z}.
\end{equation}

In terms of any complete orthonormal set, $v_h$ we can write the evaluational functional as,

\begin{equation}
    \textbf{e}_a = \lim_{k\to \infty} \sum_{h=1}^k \braket{v_h|\textbf{e}_a} v_h = \lim_{k\to \infty} \sum_{h=1}^k {v^*_h(a)} v_h.
\end{equation}

We know that strong convergence (a proven property of the SB space) implies pointwise converge which yields,

\begin{equation}
    \textbf{e}_a(z) = \sum_{h} {v^*_h(a)} v_h(z),
\end{equation}

regardless of which orthonormal basis $v_h$ we choose. If we specify a specific basis, namely,

\begin{equation}
    u_{[m]}(z) = \prod_k \frac{z_k^{m_k}}{\sqrt{m_k!}},
\end{equation}

where $m_k$ is a monotone increasing sequence of integers, we find, 

\begin{equation}
    \textbf{e}_a(z) = \sum_{m} \prod_k \frac{({a^*_k} z_k)^{m_k}}{m_k!} = e^{{a^*}\cdot z},
\end{equation}

as the Gaussian reproducing kernel. It is well known that the Gaussian kernel, expanded into a power series in the inner product of our data space, has positive coefficients. As such, the SB space is an RKHS with a universal kernel. This is useful as Gaussian kernels are the foundation of classical machine learning and through the SB space, our quantum kernel has direct access to this kernel \cite{Bargmann_1961, paulsen_raghupathi_2016}.

\section{An integration required to calculate the closed form of CV kernels}

In this section we  will calculate the integral 
\eqn{
    I_n(a,b) \ce \intall \ddd x\ \exp\lb(-ax^2+bx\rb) x^n
    \label{eq:Idef}
}
for $a,b\in\bb{C}$ with $\Re(a)>0$ and $n\in\bb{N}_0$.
Since CV quantum states, when represented as holomorphic functions, are always a product of a Gaussian and a polynomial, the evaluation of any CV quantum kernel will require the evaluation of integrals of the form of Eq.\ \eqref{eq:Idef}.

\subsection{Explicit calculation}\label{sec:InAB}

The explicit evaluation of Eq.\ \eqref{eq:Idef} is

\aln{
	I_n(a,b) &\ce \intall \ddd x\ \exp\lb(-ax^2+bx\rb) x^n \nn\\
	&= \frac{1}{a^{(n+1)/2}}\bigg[\frac{1+(-1)^n}{2}\Gamma\lb(\frac{1}{2}+\frac{n}{2}\rb)\ \hypg{1}{F}{1}\lb(\frac{1}{2}+\frac{n}{2};\ \frac{1}{2};\ \frac{b^2}{4a}\rb) \nn\\
	&\qqd + \frac{1+(-1)^{n+1}}{2}\Gamma\lb(\frac{3}{2}+\frac{n-1}{2}\rb) \frac{b}{\sqrt{a}}\ \hypg{1}{F}{1}\lb(\frac{3}{2}+\frac{n-1}{2};\ \frac{3}{2};\ \frac{b^2}{4a}\rb)\bigg]
    \label{eq:Ibasic}
}
where $\hypg{1}{F}{1}$ is the confluent hypergeometric function, which have the property that for $n\in\bb{N}_0$:
\eqn{
    \hypg{1}{F}{1}(b+n;\ b;\ \zeta) = \ec^{\zeta}\sum_{j=0}^{n} {n\choose j} \frac{\zeta^j}{(b)_j} \label{eq:HypgPoly}
}
where $(b)_j$ is the Pochhammer symbol
\eqns{
    (b)_j \ce \frac{\Gamma(b+j)}{\Gamma(b)}. %
}
We explicitly prove Eq.\ \eqref{eq:HypgPoly} in section \ref{sec:hypg}.

By considering the cases of even $n$ and odd $n$ separately, Eq.\ \eqref{eq:HypgPoly} can be used to simply Eq.\ \eqref{eq:Ibasic} and write it as a product of a Gaussian and a polynomial.

This will use of the fact that for $n\in\bb{N}_0$
\eqn{
    \Gamma\lb(\frac{1}{2}+n\rb) = \frac{\sqrt{\pi}}{2^{2n}}\frac{(2n)!}{n!}.
}

When $n$ is even:
\aln{
	I_n(a,b) &= \frac{1}{a^{(n+1)/2}} \Gamma\lb(\frac{1}{2}+\frac{n}{2}\rb)\ \hypg{1}{F}{1}\lb(\frac{1}{2}+\frac{n}{2};\ \frac{1}{2};\ \frac{b^2}{4a}\rb) \nn\\
	&= \frac{1}{a^{(n+1)/2}} \Gamma\lb(\frac{1}{2}+\frac{n}{2}\rb) \exp\lb(\frac{b^2}{4a}\rb) \sum_{j=0}^{n/2} {n/2 \choose j} \frac{1}{2^{2j}(1/2)_j} \lb(\frac{b}{\sqrt{a}}\rb)^{2j} \nn\\
	&= \frac{1}{a^{(n+1)/2}} \Gamma\lb(\frac{1}{2}+\frac{n}{2}\rb) \exp\lb(\frac{b^2}{4a}\rb) \sum_{j=0}^{n/2} {n/2 \choose j} \frac{1}{2^{2j}} \frac{\Gamma(1/2)}{\Gamma(1/2+j)} \lb(\frac{b}{\sqrt{a}}\rb)^{2j} \nn\\
        &= \frac{\sqrt{\pi}}{a^{(n+1)/2}} \exp\lb(\frac{b^2}{4a}\rb) \sum_{j=0}^{n/2} \frac{1}{2^n} \frac{n!}{(n/2-j)!(2j)!} \lb(\frac{b}{\sqrt{a}}\rb)^{2j} \nn\\
	&= \frac{\sqrt{\pi}}{a^{(n+1)/2}} \exp\lb(\frac{b^2}{4a}\rb) \sum_{k=0}^{n} \gamma_{n,k}^{(\tx{even})} \lb(\frac{b}{\sqrt{a}}\rb)^{k} \label{eq:Ieven}
}
where
\eqn{
	\gamma_{n,k}^{(\tx{even})} \ce \begin{dcases}
		\frac{1}{2^n} \frac{n!}{\big((n-k)/2\big)!k!}, & k \tx{ even} \\
		0, & k \tx{ odd}
	\end{dcases}
}
and when $n$ is odd:
\aln{
        I_n(a,b) &= \frac{1}{a^{(n+1)/2}} \Gamma\lb(\frac{3}{2}+\frac{n-1}{2}\rb) \frac{b}{\sqrt{a}}\ \hypg{1}{F}{1}\lb(\frac{3}{2}+\frac{n-1}{2};\ \frac{3}{2};\ \frac{b^2}{4a}\rb) \nn\\
        &= \frac{1}{a^{(n+1)/2}} \Gamma\lb(\frac{3}{2}+\frac{n-1}{2}\rb) \exp\lb(\frac{b^2}{4a}\rb) \sum_{j=0}^{(n-1)/2} {(n-1)/2 \choose j} \frac{1}{2^{2j}(3/2)_j} \lb(\frac{b}{\sqrt{a}}\rb)^{2j+1} \nn\\
        &= \frac{1}{a^{(n+1)/2}} \Gamma\lb(\frac{1}{2}+\frac{n+1}{2}\rb) \exp\lb(\frac{b^2}{4a}\rb) \sum_{j=0}^{(n-1)/2} {(n-1)/2 \choose j} \frac{1}{2^{2j}}\frac{\Gamma(3/2)}{\Gamma\big(1/2+(j+1)\big)} \lb(\frac{b}{\sqrt{a}}\rb)^{2j+1} \nn\\
        &= \frac{\sqrt{\pi}}{a^{(n+1)/2}} \exp\lb(\frac{b^2}{4a}\rb) \sum_{j=0}^{(n-1)/2} \frac{1}{2^n} \frac{\big((n-1)/2\big)!}{\big((n-1)/2-j\big)!j!} \frac{(n+1)!}{\big((n-1)/2+1\big)!} \frac{(j+1)!}{(2j+2)!}\lb(\frac{b}{\sqrt{a}}\rb)^{2j+1} \nn\\
        &= \frac{\sqrt{\pi}}{a^{(n+1)/2}} \exp\lb(\frac{b^2}{4a}\rb) \sum_{j=0}^{(n-1)/2} \frac{1}{2^n} \frac{n!}{\big((n-2j-1)/2\big)!\big(2j+1\big)!} \lb(\frac{b}{\sqrt{a}}\rb)^{2j+1} \nn\\
        &= \frac{\sqrt{\pi}}{a^{(n+1)/2}} \exp\lb(\frac{b^2}{4a}\rb) \sum_{k=0}^{n} \gamma_{n,k}^{(\tx{odd})} \lb(\frac{b}{\sqrt{a}}\rb)^{k} \label{eq:Iodd}
}
where
\aln{
	\gamma_{n,k}^{(\tx{odd})} &\ce \begin{dcases}
		\frac{1}{2^n} \frac{n!}{\big((n-k)/2\big)!k!}, & k \tx{ odd} \\
		0, & k \tx{ even}
	\end{dcases}
}
Therefore, combining Eqns.\ \eqref{eq:Ieven} and \eqref{eq:Iodd} gives the general result:
\eqn{
	I_n(a,b) = \frac{\sqrt{\pi}}{a^{(n+1)/2}} \exp\lb(\frac{b^2}{4a}\rb) \sum_{j=0}^{n} \gamma_{n,j} \lb(\frac{b}{\sqrt{a}}\rb)^{j} \label{eq:Igen}
}
where
\eqn{
	\gamma_{n,j} \ce \begin{dcases}
		\frac{1}{2^n} \frac{n!}{\big((n-j)/2\big)!j!}, & n\equiv j\ (\tx{mod }2) \\
		0, & \tx{otherwise}
	\end{dcases}
}

\subsection{Proof the confluent hypergeometric function $\big(\hypg{1}{F}{1}(b+n;\ b;\ \zeta)\big)$ is a product of an exponential and a polynomial}\label{sec:hypg}

In this section, we explicitly prove Eq.\ \eqref{eq:HypgPoly} and show a confluent hypergeometric function of the form  $\hypg{1}{F}{1}(b+n;\ b;\ \zeta)$ can be written as a product of an exponential and a polynomial of degree $n$.

This proof requires two properties of confluent hypergeometric functions\cite[\href{https://dlmf.nist.gov/13.6.E1}{(13.6.1)},\href{https://dlmf.nist.gov/13.3.E4}{(13.3.4)}]{NIST:DLMF}
\subeqn{
\aln{
    \hypg{1}{F}{1}(b;\ b;\ \zeta) &= \ec^{\zeta} \label{eq:Hypgneq0}\\
    \hypg{1}{F}{1}(a;\ b;\ \zeta) &= \hypg{1}{F}{1}(a-1;\ b;\ \zeta) + \frac{\zeta}{b}\ \hypg{1}{F}{1}(a;\ b+1;\ \zeta). \label{eq:Hypgrecusion}
}
}

\textbf{Proposition}:
For $n\in\bb{N}_0$, the confluent hypergeometric function
\eqn{
    \hypg{1}{F}{1}(b+n;\ b;\ \zeta) = \ec^{\zeta}\sum_{j=0}^{n} {n\choose j} \frac{\zeta^j}{(b)_j} \label{eq:HypgPolyProp}
}
where $(b)_j$ is the Pochhammer symbol.

\textbf{Proof} (by induction):

The base case for $n=0$ is given in Eq.~\eqref{eq:Hypgneq0}.

Assume that Eq.~\eqref{eq:HypgPolyProp} holds for $n$ and consider the $n+1$ case
\aln{
    \hypg{1}{F}{1}(b+n+1;\ b;\ \zeta) &= \hypg{1}{F}{1}(b+n;\ b;\ \zeta) + \frac{\zeta}{b}\ \hypg{1}{F}{1}(b+n+1;\ b+1;\ \zeta) \nn\\
    &= \ec^{\zeta}\sum_{j=0}^{n} {n\choose j} \frac{\zeta^j}{(b)_j} + \frac{\zeta}{b}\lb(\ec^{\zeta}\sum_{j=0}^{n} {n\choose j} \frac{\zeta^j}{(b+1)_j}\rb) \nn\\
    &= \ec^{\zeta} \lb(\sum_{j=0}^{n} {n\choose j} \frac{\zeta^j}{(b)_j} + \sum_{j=0}^{n} {n\choose j} \frac{\Gamma(b+1)}{\Gamma(b+j+1)} \frac{\zeta^{j+1}}{b}\rb) \nn\\
    &= \ec^{\zeta} \lb(\sum_{j=0}^{n} {n\choose j} \frac{\zeta^j}{(b)_j} + \sum_{k=1}^{n+1} {n\choose k-1} \frac{\Gamma(b)}{\Gamma(b+k)} \zeta^{k}\rb) \nn\\
    &= \ec^{\zeta} \lb(1+ \sum_{j=1}^{n} {n\choose j} \frac{\zeta^j}{(b)_j} + \sum_{j=1}^{n} {n\choose j-1} \frac{\zeta^{j}}{(b)_j} + \frac{\zeta^{n+1}}{(b)_n}\rb) \nn\\
    &= \ec^{\zeta} \lb(1+ \sum_{j=1}^{n} {n+1 \choose j} \frac{\zeta^j}{(b)_j} + \frac{\zeta^{n+1}}{(b)_n}\rb) \quad (\tx{Pascal's identity}) \nn\\
    &= \ec^{\zeta} \sum_{j=0}^{n+1} {n+1 \choose j} \frac{\zeta^j}{(b)_j}.
}
Therefore if Eq.~\eqref{eq:HypgPolyProp} holds for $n$ then it holds for $n+1$.

By the induction rule, Eq.~\eqref{eq:HypgPolyProp} holds for all $n\in\bb{N}_0$. $\square$

\section{Explicit calculation of the general $m$-mode CV kernel}\label{sec:MMKernel}

In this section, we provide the details of the calculation of the general $m$-mode kernel (Eq.\ \eqref{eq:MMkernel}).

The first step in calculating the $m$-mode kernel is to rewrite Eq.\ \eqref{eq:MMkernelstartstart} to become a series of nested integrals each of the form of Eq.\ \eqref{eq:Igen}.

Starting from the general $m$-mode inner product, 
\aln{
	\Bk{F_{\txx_1}^{\star}|F_{\txx_2}^{\star}}_{\tx{SB}} &= \frac{1}{\pi^m} \int_{\bd{z}\in\bb{C}^m}\dd^{2m}z\ \ec^{-\Abs{\bd{z}}^2} F_{\txx_1}(\bd{z})^* F_{\txx_2}(\bd{z}) \nn\\
        &= \frac{1}{\pi^m} \int_{\bd{z}\in\bb{C}^m}\dd^{2m}z\ \ec^{-\Abs{\bd{z}}^2} \exp\lb(-\frac{1}{2} \bd{z}^{\ct} \bd{A}(\txx_1)^* \bd{z}^* + \bd{B}(\txx_1)^{\ct} \bd{z}^* + C(\txx_1)^*\rb) \nn\\
	&\qqd \times \lb(\sum_{\substack{i_1,\dotsc,i_m\ge0\\ i_1+\dotsb+i_m\le n}} \beta_{\bd{i}}(\txx_1)^* (z_1^*)^{i_1} \dotsb (z_n^*)^{i_n}\rb) \nn\\
	&\qqd \times \exp\lb(-\frac{1}{2} \bd{z}^{\trps} \bd{A}(\txx_2) \bd{z} + \bd{B}(\txx_2)^{\trps} \bd{z} + C(\txx_2)\rb) \lb(\sum_{\substack{j_1\dotsc,j_m\ge0\\ j_1+\dotsb+j_m\le n}} \beta_{\bd{j}}(\txx_2) z_1^{j_1} \dotsb z_n^{j_n}\rb) \nn\\
    &= \frac{1}{\pi^m} \int_{\bd{z}\in\bb{C}^m} \ddd^{2m}\ \exp\bigg(-\Abs{\bd{z}}^2 -\frac{1}{2} \bd{z}^{\ct} \bd{A}(\txx_1)^* \bd{z}^* -\frac{1}{2} \bd{z}^{\trps} \bd{A}(\txx_2) \bd{z} \nn\\
    &\qqqqd + \bd{B}(\txx_1)^{\ct} \bd{z}^* + \bd{B}(\txx_2)^{\trps} \bd{z} + C(\txx_1)^* + C(\txx_2)\bigg) \nn\\
    &\qqd \times \lb(\sum_{\substack{i_1,\dotsc,i_m\ge0\\ i_1+\dotsb+i_m\le n}}\ \sum_{\substack{j_1,\dotsc,j_m\ge0\\ j_1+\dotsb+j_m\le n}} \beta_{\bd{i}}(\txx_1)^* \beta_{\bd{j}}(\txx_2) (z_1^*)^{i_1} z_1^{j_1} \dotsb (z_n^*)^{i_n} z_n^{j_n}\rb).
    \label{eq:MMkernelmid}
}

Unsurprisingly the integrand is also of the form
\eqn{
    \exp\big[Q(\bd{z})\big] \times P(\bd{z})
}
i.e.\ the product of a Gaussian and a polynomial.

We will set $z_j=x_{2j-1}+\iu x_{2j}$ for each $j\in[1,m]$. After some algebra, which we detail below, this substitution will allow for Eq.\ \eqref{eq:MMkernelmid} to be written as a nested set of integrals, each of the form of Eq.\ \eqref{eq:Igen}.

First simplify the Gaussian part of the integrand:
\aln{
    Q(\bd{z}) &= \Big(C(\txx_1)^*+C(\txx_2)\Big) + \sum_{j=1}^{m} -(x_{2j-1}^2+x_{2j}^2) - \frac{A_{j,j}(\txx_1)^*}{2} (x_{2j-1}-\iu x_{2j})^2 - \frac{A_{j,j}(\txx_2)}{2} (x_{2j-1}+\iu x_{2j})^2 \nn\\
    &\qd + (x_{2j-1}-\iu x_{2j})\lb(B_j(\txx_1)^* - \sum_{k=j+1}^{m} \frac{A_{j,k}(\txx_1)^*+A_{k,j}(\txx_1)^*}{2} (x_{2j-1}-\iu x_{2j})\rb) \nn\\
    &\qd + (x_{2j-1}+\iu x_{2j})\lb(B_j(\txx_2) - \sum_{k=j+1}^{m} \frac{A_{j,k}(\txx_2)+A_{k,j}(\txx_2)}{2} (x_{2j-1}+\iu x_{2j})\rb) \nn\\
    &= \Big(C(\txx_1)^*+C(\txx_2)\Big) + \sum_{j=1}^{m} -\lb(1+\frac{A_{j,j}(\txx_1)^*}{2}+\frac{A_{j,j}(\txx_2)}{2}\rb)x_{2j-1}^2 - \lb(1-\frac{A_{j,j}(\txx_1)^*}{2}-\frac{A_{j,j}(\txx_2)}{2}\rb)x_{2j}^2 \nn\\
    &\qd + x_{2j-1} \Bigg[\Big(B_j(\txx_1)^*+B_j(\txx_2)\Big) - \iu\Big(A_{j,j}(\txx_1)^*-A_{j,j}(\txx_2)\Big)x_{2j} \nn\\
    &\qqd - \frac{1}{2}\sum_{k=j+1}^{m} \Big(A_{j,k}(\txx_1)^*+A_{k,j}(\txx_1)^*+A_{j,k}(\txx_2)+A_{k,j}(\txx_2)\Big)x_{2k-1} \nn\\
    &\qqqd - \iu\Big(A_{j,k}(\txx_1)^*+A_{k,j}(\txx_1)^*-A_{j,k}(\txx_2)-A_{k,j}(\txx_2)\Big)x_{2k}\Bigg] \nn\\
    &\qd - \iu x_{2j} \Bigg[\Big(B_j(\txx_1)^*-B_j(\txx_2)\Big) \nn\\
    &\qqd - \frac{1}{2}\sum_{k=j+1}^{m} \Big(A_{j,k}(\txx_1)^*+A_{k,j}(\txx_1)^*-A_{j,k}(\txx_2)-A_{k,j}(\txx_2)\Big)x_{2k-1} \nn\\
    &\qqqd - \iu\Big(A_{j,k}(\txx_1)+A_{k,j}(\txx_1)^*+A_{j,k}(\txx_2)+A_{k,j}(\txx_2)\Big)x_{2k}\Bigg] \nn\\
    &= \Big(C(\txx_1)^*+C(\txx_2)\Big) + \sum_{j=1}^{2m} -a_{0,j}x_j^2 + x_j\lb(b_{0,j} + \sum_{k=j+1}^{2m}d_{0,j,k}x_{k}\rb)
}
where
for odd $j$
\aln{
        a_{0,j} &\ce 1+\frac{A_{(j+1)/2,(j+1)/2}(\txx_1)^*}{2}+\frac{A_{(j+1)/2,(j+1)/2}(\txx_2)}{2} \nn\\
	b_{0,j} &\ce B_{(j+1)/2}(\txx_1)^* + B_{(j+1)/2}(\txx_2) \nn\\
	d_{0,j,k} &\ce \begin{dcases}
		-\iu\Big(A_{(j+1)/2,(j+1)/2}(\txx_1)^*-A_{(j+1)/2,(j+1)/2}(\txx_2)\Big), &k=j+1 \\
            \begin{multlined}
                -\frac{1}{2}\Big(A_{(j+1)/2,(k+1)/2}(\txx_1)^*+A_{(k+1)/2,(j+1)/2}(\txx_1)^* \\
                + A_{(j+1)/2,(k+1)/2}(\txx_2)+A_{(k+1)/2,(j+1)/2}(\txx_2)\Big),
            \end{multlined} &k\ge j+2,\ k\ \tx{odd} \\ 
		\begin{multlined}
                \frac{\iu}{2} \Big(A_{(j+1)/2,k/2}(\txx_1)^* + A_{k/2,(j+1)/2}(\txx_1)^* \\
                - A_{(j+1)/2,k/2}(\txx_2)-A_{k/2,(j+1)/2}(\txx_2)\Big),
            \end{multlined} &k\ge j+3,\ k\ \tx{even}
	\end{dcases}
        \label{eq:abdStartOdd}
}

and  for even $j$
\aln{
	a_{0,j} &\ce 1-\frac{A_{j/2,j/2}(\txx_1)^*}{2}-\frac{A_{j/2,j/2}(\txx_2)}{2} \nn\\
	b_{0,j} &\ce -\iu\Big(B_{j/2}(\txx_1)^* - B_{j/2}(\txx_2)\Big) \nn\\
	d_{0,j,k} &\ce \begin{dcases}
            \begin{multlined}
                \frac{\iu}{2} \Big(A_{j/2,(k+1)/2}(\txx_1)^*+A_{(k+1)/2,j/2}(\txx_1)^* \\
                - A_{j/2,(k+1)/2}(\txx_2)-A_{(k+1)/2,j/2}(\txx_2)\Big),
            \end{multlined} &k\ge j+1,\ k\ \tx{odd} \\
		\begin{multlined}
		    \frac{1}{2} \Big(A_{j/2,k/2}(\txx_1)^*+A_{k/2,j/2}(\txx_1)^* \\
                + A_{j/2,k/2}(\txx_2)+A_{k/2,j/2}(\txx_2)\Big),
            \end{multlined} &k\ge j+2,\ k\ \tx{even}.
	\end{dcases}
        \label{eq:abdStartEven}
}

Next, simplify the polynomial part of the integrand:
\aln{
    P(\bd{z}) &= \sum_{\substack{i_1,\dotsc,i_m\ge0\\ i_1+\dotsb+i_m\le n}}\ \sum_{\substack{j_1,\dotsc,j_m\ge0\\ j_1+\dotsb+j_m\le n}} \beta_{\bd{i}}(\txx_1)^* \beta_{\bd{j}}(\txx_2) \prod_{k=1}^{m} \lb(x_{2k-1}-\iu x_{2k}\rb)^{i_k} \lb(x_{2k-1}+\iu x_{2k}\rb)^{j_k} \nn\\
    &= \sum_{\substack{i_1,\dotsc,i_m\ge0\\ i_1+\dotsb+i_m\le n}}\ \sum_{\substack{j_1,\dotsc,j_m\ge0\\ j_1+\dotsb+j_m\le n}} \beta_{\bd{i}}(\txx_1)^* \beta_{\bd{j}}(\txx_2) \nn\\
    &\qd \times \prod_{k=1}^{m} \lb(\sum_{p_k=0}^{i_k} \sum_{\qew_k=0}^{j_k} {i_k \choose p_k} {j_k \choose \qew_k} (-\iu)^{p_k} (\iu)^{\qew_k} x_{2k-1}^{i_k+j_k-p_k-\qew_k} x_{2k}^{p_k+\qew_k}\rb) \nn\\
    &= \sum_{\substack{i_1,\dotsc,i_m\ge0\\ i_1+\dotsb+i_m\le n}}\ \sum_{\substack{j_1,\dotsc,j_m\ge0\\ j_1+\dotsb+j_m\le n}} \beta_{\bd{i}}(\txx_1)^* \beta_{\bd{j}}(\txx_2) \nn\\
    &\qd \times \sum_{p_1=0}^{i_1} \sum_{\qew_1=0}^{j_1} \sum_{p_2=0}^{i_2} \sum_{\qew_2=0}^{j_2} \dotsb \sum_{p_m=0}^{i_m} \sum_{\qew_m=0}^{j_m} \prod_{k=1}^{m} {i_k \choose p_k} {j_k \choose \qew_k} (-\iu)^{p_k} (\iu)^{\qew_k} x_{2k-1}^{i_k+j_k-p_k-\qew_k} x_{2k}^{p_k+\qew_k} \nn\\
    &= \sum_{\substack{i_1,\dotsc,i_m\ge0\\ i_1+\dotsb+i_m\le n}}\ \sum_{\substack{j_1,\dotsc,j_m\ge0\\ j_1+\dotsb+j_m\le n}} \beta_{\bd{i}}(\txx_1)^* \beta_{\bd{j}}(\txx_2) \sum_{\bd{p}=\bd{0}}^{\bd{i}}  \sum_{\bd{\qew}=\bd{0}}^{\bd{j}} g(\bd{i},\bd{j},\bd{p},\bd{\qew}) \prod_{k=1}^{m} x_{2k-1}^{i_k+j_k-p_k-\qew_k} x_{2k}^{p_k+\qew_k} \nn\\
    &= \sum_{\substack{i_1,\dotsc,i_m\ge0\\ i_1+\dotsb+i_m\le n}}\ \sum_{\substack{j_1,\dotsc,j_m\ge0\\ j_1+\dotsb+j_m\le n}} \beta_{\bd{i}}(\txx_1)^* \beta_{\bd{j}}(\txx_2) \sum_{\bd{p}=\bd{0}}^{\bd{i}}  \sum_{\bd{\qew}=\bd{0}}^{\bd{j}} g(\bd{i},\bd{j},\bd{p},\bd{\qew}) \prod_{k=1}^{2m} x_k^{r_{0,k}}
}
where
\aln{
    \sum_{\bd{p}=\bd{0}}^{\bd{i}} &= \sum_{p_1=0}^{i_1} \sum_{p_2=0}^{i_2} \dotsb \sum_{p_m=0}^{i_m} \nn\\
    \sum_{\bd{\qew}=\bd{0}}^{\bd{j}} &= \sum_{\qew_1=0}^{j_1} \sum_{\qew_2=0}^{j_2} \dotsb \sum_{\qew_m=0}^{j_m} \nn\\
    g(\bd{i},\bd{j},\bd{p},\bd{\qew}) &\ce \prod_{k=1}^{m} {i_k \choose p_k} {j_k \choose \qew_k} (-\iu)^{p_k} (\iu)^{\qew_k} \nn\\ 
    r_{0,k} &\ce \pw{
		i_{(k+1)/2} + j_{(k+1)/2} - p_{(k+1)/2} - \qew_{(k+1)/2}, &k\ \tx{odd} \\
		p_{k/2}+\qew_{k/2}, & k\ \tx{even}.
    }
    \label{eq:seeds}
}
Therefore, the $m$-mode inner product can be written as:
\aln{
    \Bk{F_{\txx_1}^{\star}|F_{\txx_2}^{\star}}_{\tx{SB}} &= \frac{\ec^{C(\txx_1)^*+C(\txx_2)}}{\pi^m} \sum_{\substack{i_1,\dotsc,i_m\ge0\\ i_1+\dotsb+i_m\le n}}\ \sum_{\substack{j_1,\dotsc,j_m\ge0\\ j_1+\dotsb+j_m\le n}} \beta_{\bd{i}}(\txx_1)^* \beta_{\bd{j}}(\txx_2) \sum_{\bd{p}=\bd{0}}^{\bd{i}}  \sum_{\bd{\qew}=\bd{0}}^{\bd{j}} g(\bd{i},\bd{j},\bd{p},\bd{\qew}) \nn\\
	&\qd \times \int_{\bd{x}\in\bb{R}^{2m}} \dd^{2m}x\  \exp\lb[\sum_{j=1}^{2m} -a_{0,j}x_j^2 + x_j\lb(b_{0,j} + \sum_{k=j+1}^{2m}d_{0,j,k}x_{k}\rb)\rb] \prod_{k=1}^{2m} x_k^{r_{0,k}}
}
which is in the required form.

Now we can carry out the integrations starting with the $x_1$ integral, which is of the form Eq.\ \eqref{eq:Igen}:
\aln{
    \Bk{F_{\txx_1}^{\star}|F_{\txx_2}^{\star}}_{\tx{SB}} &= \frac{\ec^{C(\txx_1)^*+C(\txx_2)}}{\pi^m} \sum_{\substack{i_1,\dotsc,i_m\ge0\\ i_1+\dotsb+i_m\le n}}\ \sum_{\substack{j_1,\dotsc,j_m\ge0\\ j_1+\dotsb+j_m\le n}} \beta_{\bd{i}}(\txx_1)^* \beta_{\bd{j}}(\txx_2) \sum_{\bd{p}=\bd{0}}^{\bd{i}}  \sum_{\bd{\qew}=\bd{0}}^{\bd{j}} g(\bd{i},\bd{j},\bd{p},\bd{\qew})\nn\\
    &\qd \times \intall \ddd x_{2m}\ \intall \ddd x_{2m-1}\ \dotsb \intall \ddd x_2\ \nn\\
    &\qqd \times \vast\{\intall \ddd x_1\ \exp\lb[-a_{0,1}x_1^2 + x_1\lb(b_{0,1} + \sum_{k=2}^{2m} d_{0,1,k}x_{k}\rb)\rb] x_1^{r_{0,1}} \vast\} \nn\\
    &\qqd \times \exp\lb[\sum_{j=2}^{2m} -a_{0,j}x_j^2 + x_j\lb(b_{0,j} + \sum_{k=j+1}^{2m} d_{0,j,k}x_{k}\rb)\rb] \prod_{k=2}^{2m} x_k^{r_{0,k}} \nn\\
    &= \frac{\ec^{C(\txx_1)^*+C(\txx_2)}}{\pi^m} \sum_{\substack{i_1,\dotsc,i_m\ge0\\ i_1+\dotsb+i_m\le n}}\ \sum_{\substack{j_1,\dotsc,j_m\ge0\\ j_1+\dotsb+j_m\le n}} \beta_{\bd{i}}(\txx_1)^* \beta_{\bd{j}}(\txx_2) \sum_{\bd{p}=\bd{0}}^{\bd{i}}  \sum_{\bd{\qew}=\bd{0}}^{\bd{j}} g(\bd{i},\bd{j},\bd{p},\bd{\qew})\nn\\
    &\qd \times \intall \ddd x_{2m}\ \intall \ddd x_{2m-1}\ \dotsb \intall \ddd x_2\ \nn\\
    &\qqd \times \Vast\{\frac{\sqrt{\pi}}{a_{0,1}^{(r_{0,1}+1)/2}} \exp\lb[\frac{1}{4a_{0,1}}\lb(b_{0,1} + \sum_{k=2}^{2m} d_{0,1,k}x_{k}\rb)^2\rb] \nn\\
    &\qqqd \times \sum_{s_1=0}^{r_{0,1}} \gamma_{r_{0,1},s_1} \lb[\frac{1}{\sqrt{a_{0,1}}}\lb(b_{0,1} + \sum_{k=2}^{2m} d_{0,1,k}x_{k}\rb)\rb]^{s_1} \Vast\} \nn\\
    &\qqd \times \exp\lb[\sum_{j=2}^{2m} -a_{0,j}x_j^2 + x_j\lb(b_{0,j} + \sum_{k=j+1}^{2m} d_{0,j,k}x_{k}\rb)\rb] \prod_{k=2}^{2m} x_k^{r_{0,k}} \nn
}
\aln{
    \phantom{\Bk{F_{\txx_1}^{\star}|F_{\txx_2}^{\star}}_{\tx{SB}}} &= \frac{\ec^{C(\txx_1)^*+C(\txx_2)}}{\pi^{(2m-1)/2}} \sum_{\substack{i_1,\dotsc,i_m\ge0\\ i_1+\dotsb+i_m\le n}}\ \sum_{\substack{j_1,\dotsc,j_m\ge0\\ j_1+\dotsb+j_m\le n}} \beta_{\bd{i}}(\txx_1)^* \beta_{\bd{j}}(\txx_2) \sum_{\bd{p}=\bd{0}}^{\bd{i}}  \sum_{\bd{\qew}=\bd{0}}^{\bd{j}} g(\bd{i},\bd{j},\bd{p},\bd{\qew}) \nn\\
    &\qd \times \intall \ddd x_{2m}\ \intall \ddd x_{2m-1}\ \dotsb \intall \ddd x_2\ \exp\lb[\frac{1}{4a_{0,1}}\lb(b_{0,1} + \sum_{k=2}^{2m} d_{0,1,k}x_{k}\rb)^2\rb] \nn\\
    &\qqd \times \exp\lb[\sum_{j=2}^{2m} -a_{0,j}x_j^2 + x_j\lb(b_{0,j} + \sum_{k=j+1}^{2m} d_{0,j,k}x_{k}\rb)\rb] \nn\\
    &\qqd \times \vast[\frac{1}{a_{0,1}^{(r_{0,1}+1)/2}} \sum_{s_1=0}^{r_{0,1}} \gamma_{r_{0,1},s_1} \lb[\frac{1}{\sqrt{a_{0,1}}}\lb(b_{0,1} + \sum_{k=2}^{2m} d_{0,1,k}x_{k}\rb)\rb]^{s_1}\vast] \prod_{k=2}^{2m} x_k^{r_{0,k}}.
}

Next, the goal is to simplify the expression so that the $x_2$ integral is of the form of Eq.\ \eqref{eq:Igen}.

Start by simplifying the exponential part:
\aln{
    &\frac{1}{4a_{0,1}}\lb(b_{0,1} + \sum_{j=2}^{2m} d_{0,1,j}x_{j}\rb)^2 + \sum_{j=2}^{2m} -a_{0,j}x_j^2 + x_j\lb(b_{0,j} + \sum_{k=j+1}^{2m} d_{0,j,k}x_{k}\rb) \nn\\
    &= \frac{b_{0,1}^2}{4a_{0,1}} + \sum_{j=2}^{2m} -\lb(a_{0,j}-\frac{d_{0,1,j}^2}{4a_{0,1}}\rb) x_j^2 + x_j\Bigg[\lb(b_{0,j}+\frac{b_{0,1}d_{0,1,j}}{2a_{0,1}}\rb) \nn\\
    &\qd + \sum_{k=j+1}^{2m} \lb(d_{0,j,k} + \frac{d_{0,1,j}d_{0,1,k}}{2a_{0,1}}\rb) x_k\Bigg] \nn\\
    &= \frac{b_{0,1}^2}{4a_{0,1}} + \sum_{j=2}^{2m} -a_{1,j}x_j^2 + x_j\lb(b_{1,j} + \sum_{k=j+1}^{2m} d_{1,j,k}x_{k}\rb)
}
where
\aln{
    a_{1,j} &\ce a_{0,j}-\frac{d_{0,1,j}^2}{4a_{0,1}} \nn\\
    b_{1,j} &\ce b_{0,j}+\frac{b_{0,1}d_{0,1,j}}{2a_{0,1}} \nn\\
    d_{1,j,k} &\ce d_{0,j,k} + \frac{d_{0,1,j}d_{0,1,k}}{2a_{0,1}}.
}

Then use the multinomial expansion,
\eqn{
    \lb(x_1+x_2+\dotsb+x_m\rb)^n = \sum_{\substack{i_1,i_2,\dotsc,i_m\ge0\\ i_1+i_2+\dotsb+i_m=n}} \frac{n!}{i_1!i_2!\dotsb i_m!}\ x_1^{i_1} x_2^{i_2} \dotsb x_m^{i_m}
}
to simplify the polynomial part:
\aln{
    &\lb[\frac{1}{a_{0,1}^{(r_{0,1}+1)/2}} \sum_{s_1=0}^{r_{0,1}} \gamma_{r_{0,1},s_1} \lb(\frac{1}{\sqrt{a_{0,1}}}\rb)^{s_1} \lb(b_{0,1} + \sum_{j=2}^{2m} d_{0,1,j}x_{j}\rb)^{s_1}\rb] \prod_{k=2}^{2m} x_k^{r_{0,k}} \nn\\
    &\qd = \lb[\sum_{s_1=0}^{r_{0,1}} \frac{\gamma_{r_{0,1},s_1}}{a_{0,1}^{(r_{0,1}+s_1+1)/2}} \sum_{t_1=0}^{s_1} {s_1 \choose t_1} b_{0,1}^{s_1-t_1} \lb(\sum_{j=2}^{2m} d_{0,1,j}x_{j}\rb)^{t_1}\rb] \prod_{k=2}^{2m} x_k^{r_{0,k}} \nn\\
    &\qd = \sum_{s_1=0}^{r_{0,1}} \frac{\gamma_{r_{0,1},s_1}}{a_{0,1}^{(r_{0,1}+s_1+1)/2}} \sum_{t_1=0}^{s_1} {s_1 \choose t_1} b_{0,1}^{s_1-t_1} \nn\\
    &\qqd \times \lb(\sum_{\substack{u_{1,2},\dotsc,u_{1,2m}\ge0\\ u_{1,2}+\dotsb+u_{1,2m}=t_1}} \frac{t_1!}{u_{1,2}!\dotsb u_{1,2m}!} \prod_{k=2}^{2m} d_{0,1,k}^{u_{1,k}} x_{k}^{u_{1,k}} \rb) \prod_{k=2}^{2m} x_k^{r_{0,k}} \nn\\
    &\qd = \sum_{s_1=0}^{r_{0,1}} \frac{\gamma_{r_{0,1},s_1}}{a_{0,1}^{(r_{0,1}+s_1+1)/2}} \sum_{t_1=0}^{s_1} \frac{s_1!}{(s_1-t_1)!} b_{0,1}^{s_1-t_1} \sum_{\substack{u_{1,2},\dotsc,u_{1,2m}\ge0\\ u_{1,2}+\dotsb+u_{1,2m}=t_1}}\ \prod_{k=2}^{2m} \frac{d_{0,1,k}^{u_{1,k}}}{u_{1,k}!} x_{k}^{u_{1,k}+r_{0,k}} \nn\\
    &\qd = \sum_{s_1=0}^{r_{0,1}} \frac{\gamma_{r_{0,1},s_1}}{a_{0,1}^{(r_{0,1}+s_1+1)/2}} \sum_{t_1=0}^{s_1} \frac{s_1!}{(s_1-t_1)!} b_{0,1}^{s_1-t_1} \sum_{\substack{u_{1,2},\dotsc,u_{1,2m}\ge0\\ u_{1,2}+\dotsb+u_{1,2m}=t_1}}\ \prod_{k=2}^{2m} \frac{d_{0,1,k}^{u_{1,k}}}{u_{1,k}!} x_{k}^{r_{1,k}}
}
where for $k\ge2$
\aln{
	r_{1,k} &\ce r_{0,k} + u_{1,k}
}
Therefore the $m$-mode inner product becomes
\aln{
    \Bk{F_{\txx_1}^{\star}|F_{\txx_2}^{\star}}_{\tx{SB}} &= \frac{1}{\pi^{(2m-1)/2}} \exp\lb(C(\txx_1)^*+C(\txx_2) + \frac{b_{0,1}^2}{4a_{0,1}}\rb) \sum_{\substack{i_1,\dotsc,i_m\ge0\\ i_1+\dotsb+i_m\le n}}\ \sum_{\substack{j_1,\dotsc,j_m\ge0\\ j_1+\dotsb+j_m\le n}} \beta_{\bd{i}}(\txx_1)^* \beta_{\bd{j}}(\txx_2) \nn\\
    &\qd \times \sum_{\bd{p}=\bd{0}}^{\bd{i}} \sum_{\bd{\qew}=\bd{0}}^{\bd{j}} g(\bd{i},\bd{j},\bd{p},\bd{\qew}) \sum_{s_1=0}^{r_{0,1}}\ \frac{\gamma_{r_{0,1},s_1}}{a_{0,1}^{(r_{0,1}+s_1+1)/2}} \sum_{t_1=0}^{s_1} \frac{s_1!}{(s_1-t_1)!} b_{0,1}^{s_1-t_1} \nn\\
    &\qqd \times \sum_{\substack{u_{1,2},\dotsc,u_{1,2m}\ge0\\ u_{1,2}+\dotsb+u_{1,2m}=t_1}} \intall \ddd x_{2m}\ \dotsb \intall \ddd x_2\ \nn\\
    &\qqqd\times \exp\lb[\sum_{j=2}^{2m} -a_{1,j}x_j^2 + x_j\lb(b_{1,j} + \sum_{k=j+1}^{2m} d_{1,j,k}x_{k}\rb)\rb] \prod_{k=2}^{2m} \frac{d_{0,1,k}^{u_{1,k}}}{u_{1,k}!} x_{k}^{r_{1,k}} \nn\\
    &= \frac{1}{\pi^{(2m-1)/2}} \exp\lb(C(\txx_1)^*+C(\txx_2) + \frac{b_{0,1}^2}{4a_{0,1}}\rb) \sum_{\substack{i_1,\dotsc,i_m\ge0\\ i_1+\dotsb+i_m\le n}}\ \sum_{\substack{j_1,\dotsc,j_m\ge0\\ j_1+\dotsb+j_m\le n}} \beta_{\bd{i}}(\txx_1)^* \beta_{\bd{j}}(\txx_2) \nn\\
    &\qd \times \sum_{\bd{p}=\bd{0}}^{\bd{i}} \sum_{\bd{\qew}=\bd{0}}^{\bd{j}} g(\bd{i},\bd{j},\bd{p},\bd{\qew}) \sum_{s_1=0}^{r_{0,1}}\ \frac{\gamma_{r_{0,1},s_1}}{a_{0,1}^{(r_{0,1}+s_1+1)/2}} \sum_{t_1=0}^{s_1} \frac{s_1!}{(s_1-t_1)!} b_{0,1}^{s_1-t_1} \nn\\
    &\qqd \times \sum_{\substack{u_{1,2},\dotsc,u_{1,2m}\ge0\\ u_{1,2}+\dotsb+u_{1,2m}=t_1}} \lb(\prod_{k=2}^{2m} \frac{d_{0,1,k}^{u_{1,k}}}{u_{1,k}!}\rb) \intall \ddd x_{2m}\ \dotsb \intall \ddd x_3\ \nn\\
    &\qqqd \times \vast\{\intall \ddd x_2\ \exp\lb[-a_{1,2}x_2^2 + x_2\lb(b_{1,2} + \sum_{k=3}^{2m} d_{1,2,k}x_{k}\rb)\rb] x_2^{r_{1,2}} \vast\} \nn\\
    &\qqqd \times \exp\lb[\sum_{j=3}^{2m} -a_{1,j}x_j^2 + x_j\lb(b_{1,j} + \sum_{k=j+1}^{2m} d_{1,j,k}x_{k}\rb)\rb] \prod_{k=3}^{2m} x_{k}^{r_{1,k}} \nn
}
\aln{
    \phantom{\Bk{F_{\txx_1}^{\star}|F_{\txx_2}^{\star}}_{\tx{SB}}} &= \frac{1}{\pi^{(2m-1)/2}} \exp\lb(C(\txx_1)^*+C(\txx_2) + \frac{b_{0,1}^2}{4a_{0,1}}\rb) \sum_{\substack{i_1,\dotsc,i_m\ge0\\ i_1+\dotsb+i_m\le n}}\ \sum_{\substack{j_1,\dotsc,j_m\ge0\\ j_1+\dotsb+j_m\le n}} \beta_{\bd{i}}(\txx_1)^* \beta_{\bd{j}}(\txx_2) \nn\\
    &\qd \times \sum_{\bd{p}=\bd{0}}^{\bd{i}} \sum_{\bd{\qew}=\bd{0}}^{\bd{j}} g(\bd{i},\bd{j},\bd{p},\bd{\qew}) \sum_{s_1=0}^{r_{0,1}}\ \frac{\gamma_{r_{0,1},s_1}}{a_{0,1}^{(r_{0,1}+s_1+1)/2}} \sum_{t_1=0}^{s_1} \frac{s_1!}{(s_1-t_1)!} b_{0,1}^{s_1-t_1} \nn\\
    &\qqd \times \sum_{\substack{u_{1,2},\dotsc,u_{1,2m}\ge0\\ u_{1,2}+\dotsb+u_{1,2m}=t_1}} \lb(\prod_{k=2}^{2m} \frac{d_{0,1,k}^{u_{1,k}}}{u_{1,k}!}\rb) \intall \ddd x_{2m}\ \dotsb \intall \ddd x_3\ \nn\\
    &\qqqd \times \Vast\{\frac{\sqrt{\pi}}{a_{1,2}^{(r_{1,2}+1)/2}} \exp\lb[\frac{1}{4a_{1,2}}\lb(b_{1,2}+\sum_{k=3}^{2m} d_{1,2,k} x_k\rb)\rb]^2 \nn\\
    &\qqqqd \times \sum_{s_2=0}^{r_{1,2}} \gamma_{r_{1,2},s_2} \lb[\frac{1}{\sqrt{a_{1,2}}} 
 \lb(b_{1,2}+\sum_{k=3}^{2m} d_{1,2,k}x_k\rb)\rb]^{s_2}\Vast\} \nn\\
    &\qqqd \times \exp\lb[\sum_{j=3}^{2m} -a_{1,j}x_j^2 + x_j\lb(b_{1,j} + \sum_{k=j+1}^{2m} d_{1,j,k}x_{k}\rb)\rb] \prod_{k=3}^{2m} x_{k}^{r_{1,k}} \nn\\
    &= \frac{1}{\pi^{m-1}} \exp\lb(C(\txx_1)^*+C(\txx_2) + \frac{b_{0,1}^2}{4a_{0,1}} + \frac{b_{1,2}^2}{4a_{1,2}}\rb) \sum_{\substack{i_1,\dotsc,i_m\ge0\\ i_1+\dotsb+i_m\le n}}\ \sum_{\substack{j_1,\dotsc,j_m\ge0\\ j_1+\dotsb+j_m\le n}} \beta_{\bd{i}}(\txx_1)^* \beta_{\bd{j}}(\txx_2) \nn\\
    &\qd \times \sum_{\bd{p}=\bd{0}}^{\bd{i}} \sum_{\bd{\qew}=\bd{0}}^{\bd{j}} g(\bd{i},\bd{j},\bd{p},\bd{\qew}) \sum_{s_1=0}^{r_{0,1}}\ \frac{\gamma_{r_{0,1},s_1}}{a_{0,1}^{(r_{0,1}+s_1+1)/2}} \sum_{t_1=0}^{s_1} \frac{s_1!}{(s_1-t_1)!} b_{0,1}^{s_1-t_1} \nn\\
    &\qqd \times \sum_{\substack{u_{1,2},\dotsc,u_{1,2m}\ge0\\ u_{1,2}+\dotsb+u_{1,2m}=t_1}} \lb(\prod_{k=2}^{2m} \frac{d_{0,1,k}^{u_{1,k}}}{u_{1,k}!}\rb) \sum_{s_2=0}^{r_{1,2}}\ \frac{\gamma_{r_{1,2},s_2}}{a_{1,2}^{(r_{1,2}+s_2+1)/2}} \sum_{t_2=0}^{s_2} \frac{s_2!}{(s_2-t_2)!} b_{1,2}^{s_2-t_2} \nn\\
    &\qqqd \times \sum_{\substack{u_{2,3},\dotsc,u_{2,2m}\ge0\\ u_{2,3}+\dotsb+u_{2,2m}=t_2}} \lb(\prod_{k=3}^{2m} \frac{d_{1,2,k}^{u_{2,k}}}{u_{2,k}!}\rb)\intall \ddd x_{2m}\ \dotsb \intall \ddd x_3\ \nn\\
    &\qqqqd \times \exp\lb[\sum_{j=3}^{2m} -a_{2,j}x_j^2 + x_j\lb(b_{2,j} + \sum_{k=j+1}^{2m} d_{2,j,k}x_{k}\rb)\rb] \prod_{k=3}^{2m} x_{k}^{r_{2,k}}
}
where the last equality results from following a similar simplification as was done after the integration with respect to $x_1$ and we define:
\aln{
    a_{2,j} &\ce a_{1,j}-\frac{d_{1,2,j}^2}{4a_{1,2}} \nn\\
    b_{2,j} &\ce b_{1,j}+\frac{b_{1,2}d_{1,2,j}}{2a_{1,2}} \nn\\
    d_{2,j,k} &\ce d_{1,j,k} + \frac{d_{1,2,j}d_{1,2,k}}{2a_{1,2}} \nn\\
    r_{2,k} &\ce r_{1,k} + u_{2,k}
}

The next $(2m-2)$ integrals can be evaluated iteratively following the same steps as above which results in the following closed form of the $m$-mode inner product: 
\aln{
    \Bk{F_{\txx_1}^{\star}|F_{\txx_2}^{\star}}_{\tx{SB}} &= \exp\lb(C(\txx_1)^*+C(\txx_2) + \sum_{j=1}^{2m} \frac{b_{j-1,j}^2}{4a_{j-1,j}}\rb) \sum_{\substack{i_1,\dotsc,i_m\ge0\\ i_1+\dotsb+i_m\le n}}\ \sum_{\substack{j_1,\dotsc,j_m\ge0\\ j_1+\dotsb+j_m\le n}} \beta_{\bd{i}}(\txx_1)^* \beta_{\bd{j}}(\txx_2) \nn\\ 
    &\qd \times \sum_{\bd{p}=\bd{0}}^{\bd{i}} \sum_{\bd{\qew}=\bd{0}}^{\bd{j}} g(\bd{i},\bd{j},\bd{p},\bd{\qew}) \Vast\{\prod_{\ell=1}^{2m-1} \Vast[\sum_{s_\ell=0}^{r_{\ell-1,\ell}} \frac{\gamma_{r_{\ell-1,\ell},s_\ell}}{a_{\ell-1,\ell}^{(r_{\ell-1,\ell}+s_\ell+1)/2}} \nn\\
    &\qqd \times \sum_{t_{\ell}=0}^{s_{\ell}} \frac{s_{\ell}!}{(s_{\ell}-t_{\ell})!} b_{\ell-1,\ell}^{s_{\ell}-t_{\ell}} \sum_{\substack{u_{\ell,\ell+1},\dotsc,u_{\ell,2m}\ge0\\ u_{\ell,\ell+1}+\dotsb+u_{\ell,2m}=t_\ell}} \lb(\prod_{k=\ell+1}^{2m} \frac{d_{\ell-1,\ell,k}^{u_{\ell,k}}}{u_{\ell,k}!}\rb)\Vast] \nn\\
    &\qqqd  \times \lb[\sum_{s_{2m}=0}^{r_{2m-1,2m}} \frac{\gamma_{r_{2m-1,2m},s_{2m}}}{a_{2m-1,2m}^{(r_{2m-1,2m}+s_{2m}+1)/2}}\ b_{2m-1,2m}^{s_{2m}}\rb]\Vast\}
}
where
\aln{
    a_{i,j} &\ce a_{i-1,j}-\frac{d_{i-1,i,j}^2}{4a_{i-1,i}} \nn\\
    b_{i,j} &\ce b_{i-1,j}+\frac{b_{i-1,i}d_{i-1,i,j}}{2a_{i-1,i}} \nn\\
    d_{i,j,k} &\ce d_{i-1,j,k} + \frac{d_{i-1,i,j}d_{i-1,i,k}}{2a_{i-1,i}} \nn\\
    r_{i,k} &\ce r_{i-1,k} + u_{i,k}.
}
are defined recursively and the $\gamma_{r,s}$'s are defined in Eq.\ \eqref{eq:gammagen}.  We note that this expression the product of Gaussian and an algebraic function in the parameters of the feature map: $A_{i,j}(\txx_k)$, $B_{i}(\txx_k)$, $C(\txx_k)$, $\beta_{\bd{i}}(\txx_k)$

\section{Approximating CV kernels of infinite stellar rank}
\label{sec:infinitekernels}

In this section we will show that kernels formed from pure states of infinite stellar rank can be approximated arbitrarily well by kernels of finite stellar rank.

CV states of infinite stellar rank can be approximated arbitrarily well in trace distance by states of finite stellar rank \cite{Chabaud_2022}. That is
\eqn{
    T\big(\Kb{\psi}{\psi},\Kb{F}{F}\big) \le \epsilon
}
where we use $\Ket{\psi}$ to denote a state of infinite stellar rank and $\Ket{F}$ to denote a state of finite stellar rank. Since we are considering pure states, the trace distance can be easily expressed in terms of the inner product
\eqn{
    T\big(\Kb{\psi}{\psi},\Kb{F}{F}\big) = \sqrt{1-\Abs{\Bk{\psi|F}}^2} \le \epsilon \imp 1-\Abs{\Bk{\psi|F}}^2 \le \epsilon^2.
}
Now since
\eqn{
    \Bk{\psi|F} = \ec^{\iu\theta}\Abs{\Bk{\psi|F}}
}
where $\theta\in[0,2\pi)$ is the phase, we can define a new state with the same stellar rank as $\Ket{F}$ as
\eqn{
    \ket{\ti{F}} \ce \ec^{-\iu\theta} \Ket{F}
}
so that
\eqn{
    \braket{\psi|\ti{F}} = \ec^{-\iu\theta}\Bk{\psi|F} = \ec^{-\iu\theta}\big(\ec^{\iu\theta}\Abs{\Bk{\psi|F}}\big) = \Abs{\Bk{\psi|F}}.
    \label{eq:tiF1}
}
Also note that for any state $\Ket{\phi}$,
\eqn{
    \Abs{\braket{\phi|\ti{F}}} = \sqrt{\braket{\phi|\ti{F}}\braket{\ti{F}|\phi}} = \sqrt{e^{-\iu\theta}\Bk{\phi|F}\ec^{\iu\theta}\Bk{F|\phi}} = \sqrt{\Bk{\phi|F}\Bk{F|\phi}} = \Abs{\Bk{\phi|F}}
    \label{eq:tiF1IP}
}
and
\aln{
    \Abs{\braket{\psi-\ti{F}|\phi}} &\le \sqrt{\braket{\psi-\ti{F}|\psi-\ti{F}}\Bk{\phi|\phi}} & \tx{(Cauchy-Schwarz)} \nn\\
    &= \sqrt{\braket{\psi-\ti{F}|\psi-\ti{F}}} \nn\\
    &= \sqrt{\Bk{\psi|\psi} - \braket{\psi|\ti{F}} - \braket{\ti{F}|\psi} + \braket{\ti{F}|\ti{F}}} \nn\\
    &= \sqrt{2-2\Abs{\Bk{\psi|F}}} & \tx{(Eq.\ \eqref{eq:tiF1})} \nn\\
    &\le \sqrt{2-2\Abs{\Bk{\psi|F}}^2} & \lb(\tx{since } \Abs{z}\le1 \imp \Abs{z}\ge\Abs{z}^2\rb) \nn\\ %
    &\le \sqrt{2}\epsilon \label{eq:diff}.
}

Next stating from 
\eqn{
    T\big(\Kb{\psi_1}{\psi_1},\Kb{F_1}{F_1}\big) \le \epsilon
}
we will bound the inner product between the state of infinite stellar rank and another state of finite stellar rank.
\aln{
    \Abs{\Bk{\psi_1|F_2}} &= \Abs{\braket{\psi_1-\ti{F}_1+\ti{F}_1|F_2}} \nn\\
    &= \Abs{\braket{\psi_1-\ti{F}_1|F_2} + \braket{\ti{F}_1|F_2}} \nn\\
    &\le \Abs{\braket{\psi_1-\ti{F}_1|F_2}} + \Abs{\braket{\ti{F}_1|F_2}} \nn\\ %
    &\le \sqrt{2}\epsilon + \Abs{\Bk{F_1|F_2}}. & \tx{(Eqs.\ \eqref{eq:tiF1IP} \& \eqref{eq:diff})}
}
Therefore
\eqn{
    \Abs{\Bk{\psi_1|F_2}} - \Abs{\Bk{F_1|F_2}} \le \sqrt{2}\epsilon.
	\label{eq:part1}
}
Similarly, we can show
\aln{
    \Abs{\Bk{F_1|F_2}} &= \Abs{\braket{\ti{F}_1|F_2}} \nn\\
    &= \Abs{\braket{\ti{F}_1-\psi_1+\psi_1|F_2}} \nn\\
    &= \Abs{\braket{\ti{F}_1-\psi_1|F_2} + \Bk{\psi_1|F_2}} \nn\\
    &\le \Abs{\braket{\ti{F}_1-\psi_1|F_2}} + \Abs{\Bk{\psi_1|F_2}} \nn\\ %
    &\le \sqrt{2}\epsilon + \Abs{\Bk{\psi_1|F_2}}. & \tx{(Eq.\ \eqref{eq:diff})}
}
Therefore
\eqn{
    \Abs{\Bk{F_1|F_2}} - \Abs{\Bk{\psi_1|F_2}} \le \sqrt{2}\epsilon
    \label{eq:part2}
}
and combining Eqs.\ \eqref{eq:part1} and \eqref{eq:part2} gives
\eqn{
    \Big|\Abs{\Bk{\psi_1|F_2}} - \Abs{\Bk{F_1|F_2}}\Big| \le \sqrt{2}\epsilon.
    \label{eq:part3}
}

Similarly, for the state of infinite finite stellar rank $\Ket{\psi_2}$ that is close in trace distance to the state of finite stellar rank $\Ket{F_2}$:
\eqn{
    T\big(\Kb{\psi_2}{\psi_2},\Kb{F_2}{F_2}\big) \le \epsilon_2
    \imp \Big|\Abs{\Bk{F_1|\psi_2}} - \Abs{\Bk{F_1|F_2}}\Big| \le \sqrt{2}\epsilon_2.
    \label{eq:part4}
}

Eqs.\ \eqref{eq:part3} and \eqref{eq:part4} can be combined to bound on the inner product 
 of two states of infinite stellar rank,
\aln{
    \Abs{\Bk{\psi_1|\psi_2}} &= \Abs{\braket{\psi_1-\ti{F}_1+\ti{F}_1|\psi_2}} \nn\\
    &= \Abs{\braket{\psi_1-\ti{F}_1|\psi_2} + \braket{\ti{F}_1|\psi_2}} \nn\\
    &\le \Abs{\braket{\psi_1-\ti{F}_1|\psi_2}} + \Abs{\braket{\ti{F}_1|\psi_2}} \nn\\ %
    &\le \sqrt{2}\epsilon + \Abs{\Bk{F_1|\psi_2}} & \tx{(Eqs.\ \eqref{eq:tiF1IP} \& \eqref{eq:diff})} \nn\\
    &\le \sqrt{2}\epsilon + \Abs{\Bk{F_1|F_2} }+ \sqrt{2}\epsilon_2 & \tx{(Eq.\ \eqref{eq:part4})} \nn\\
    &= \Abs{\Bk{F_1|F_2}} + \sqrt{2}(\epsilon+\epsilon_2).
}
Therefore
\eqn{
    \Abs{\Bk{\psi_1|\psi_2}} - \Abs{\Bk{F_1|F_2}} \le 2\sqrt{2}\ \ti{\epsilon}
    \label{eq:part5}
}
where
\eqn{
    \ti{\epsilon} \ce \max(\epsilon,\epsilon_2).
}

And
\aln{
    \Abs{\Bk{F_1|F_2}} &\le \Abs{\Bk{F_1|\psi_2}} + \sqrt{2}\epsilon_2 & \tx{(Eq.\ \eqref{eq:part4})} \nn\\
    &= \Abs{\braket{\ti{F}_1|\psi_2}} + \sqrt{2}\epsilon_2 &\tx{(Eq.\ \eqref{eq:tiF1IP})} \nn\\
    &= \Abs{\braket{\ti{F}_1-\psi_1+\psi_1|\psi_2}} + \sqrt{2}\epsilon_2 \nn\\
    &= \Abs{\braket{\ti{F}_1-\psi_1|\psi_2} + \Bk{\psi_1|\psi_2}} + \sqrt{2}\epsilon_2 \nn\\
    &\le \Abs{\braket{\ti{F}_1-\psi_1|\psi_2}} + \Abs{\Bk{\psi_1|\psi_2}} + \sqrt{2}\epsilon_2 \nn\\ %
    &\le \sqrt{2}\epsilon + \Abs{\Bk{\psi_1|\psi_2}} + \sqrt{2}\epsilon_2 & \tx{(Eq.\ \eqref{eq:diff})} \nn\\
    &= \Abs{\Bk{\psi_1|\psi_2}} + \sqrt{2}(\epsilon+\epsilon_2).
}
Therefore
\eqn{
    \Abs{\Bk{F_1|F_2}} - \Abs{\Bk{\psi_1|\psi_2}} \le 2\sqrt{2}\ \ti{\epsilon}
    \label{eq:part6}
}
and combining Eqs.\ \eqref{eq:part5} and \eqref{eq:part6} gives
\eqn{
    \Big|\Abs{\Bk{\psi_1|\psi_2}} - \Abs{\Bk{F_1|F_2}}\Big| \le 2\sqrt{2}\ \ti{\epsilon}.
}

Finally, using the fact that
\eqn{
    \Abs{\Bk{\psi_1|\psi_2}}, \Abs{\Bk{F_1|F_2}} \in [0,1] \imp 0\le\Big(\Abs{\Bk{\psi_1|\psi_2}}+\Abs{\Bk{F_1|F_2}}\Big)\le2
}
it can be shown that
\aln{
    \Abs{\Bk{\psi_1|\psi_2}}^2 - \Abs{\Bk{F_1|F_2}}^2 &= \Big(\Abs{\Bk{\psi_1|\psi_2}} - \Abs{\Bk{F_1|F_2}}\Big)\Big(\Abs{\Bk{\psi_1|\psi_2}} + \Abs{\Bk{F_1|F_2}}\Big) \nn\\
    &\le 2\sqrt{2}\ \ti{\epsilon}\ \Big(\Abs{\Bk{\psi_1|\psi_2}} + \Abs{\Bk{F_1|F_2}}\Big) \nn\\
    &\le 4\sqrt{2}\ \ti{\epsilon}
}
and
\aln{
    \Abs{\Bk{F_1|F_2}}^2 - \Abs{\Bk{\psi_1|\psi_2}}^2 &= \Big(\Abs{\Bk{F_1|F_2}} - \Abs{\Bk{\psi_1|\psi_2}}\Big)\Big(\Abs{\Bk{F_1|F_2}} + \Abs{\Bk{\psi_1|\psi_2}}\Big) \nn\\
    &\le 2\sqrt{2}\ \ti{\epsilon}\ \Big(\Abs{\Bk{F_1|F_2}} + \Abs{\Bk{\psi_1|\psi_2}}\Big) \nn\\
    &\le 4\sqrt{2}\ \ti{\epsilon}
}
so
\eqn{
    \Big|\Abs{\Bk{\psi_1|\psi_2}}^2 - \Abs{\Bk{F_1|F_2}}^2\Big| \le 4\sqrt{2}\ \ti{\epsilon}.
}

In other words the kernel defined by CV states of infinite stellar rank can be approximated arbitraily well by a CV kernel of finite stellar rank
\eqn{
    \Abs{k_{\infi}(x,x') - k_n(x,x')} \le 4\sqrt{2}\ \ti{\epsilon}
}

\section{Properties of the displaced Fock state kernel}\label{sec:KDdetails}

In this section we show the details of the derivation and properties of the displaced Fock state kernel (Eq.\ \eqref{eq:Dkernel}).

\subsection{Derivation of Eq.\ \eqref{eq:Dkernel}}

The displaced Fock state inner product is calculated from Eq.\ \eqref{eq:Dkernelstart} by first setting  $z= x+\iu y$ and applying the trinomial expansion so that the $x$ and $y$ integrals can be written in the form of Eq.\ \eqref{eq:Igen}.

\aln{
    \Bk{F_{\alpha}^{\star}(z)|F_{\beta}^{\star}(z)}_{\tx{SB}} &= \frac{1}{\pi} \frac{\ec^{-(\Abs{\alpha}^2+\Abs{\beta}^2)/2}}{n!} \intall \ddd x\ \intall \ddd y\ \ec^{-(x^2+y^2)+\alpha^*(x-\iu  y)+\beta(x+\iu y)} \nn\\
    &\qd \times (x-\iu y-\alpha)^n (x+\iu y-\beta^*)^n \nn\\
    &= \frac{1}{\pi} \frac{\ec^{-(\Abs{\alpha}^2+\Abs{\beta}^2)/2}}{n!}  \intall \ddd x\ \intall \ddd y\ \ec^{-x^2+(\alpha^*+\beta) x} \ec^{-y^2-\iu(\alpha^*-\beta) y}  \nn\\
    &\qd \times \lb(\sum_{i=0}^{n} \sum_{j=0}^{n-i} \frac{n!}{i! j! (n-i-j)!} x^i (-\iu y)^j (-\alpha)^{n-i-j}\rb) \nn\\
    &\qd \times \lb(\sum_{k=0}^{n} \sum_{\ell=0}^{n-k} \frac{n!}{k! \ell! (n-k-\ell)!} x^p (\iu y)^{\ell} (-\beta^*)^{n-k-\ell}\rb) \nn\\
    &= \frac{n!}{\pi} \ec^{-(\Abs{\alpha}^2+\Abs{\beta}^2)/2} \sum_{i=0}^{n} \sum_{j=0}^{n-i} \sum_{k=0}^{n} \sum_{\ell=0}^{n-k} \frac{(-\iu)^j (-\alpha)^{n-i-j}}{i! j! (n-i-j)!} \frac{(\iu)^{\ell} (-\beta^*)^{n-k-\ell}}{k! \ell! (n-k-\ell)!} \nn\\
    &\qd \times \lb(\intall \ddd x\ \ec^{-x^2+(\alpha^*+\beta)x} x^{i+k}\rb) \lb(\intall \ddd y\ \ec^{-y^2-\iu(\alpha^*-\beta)y} y^{j+\ell}\rb) \nn\\
    &= \frac{n!}{\pi} \ec^{-(\Abs{\alpha}^2+\Abs{\beta}^2)/2} \sum_{i=0}^{n} \sum_{j=0}^{n-i} \sum_{k=0}^{n} \sum_{\ell=0}^{n-k} \frac{(-\iu)^j (-\alpha)^{n-i-j}}{i! j! (n-i-j)!} \frac{(\iu)^{\ell} (-\beta^*)^{n-k-\ell}}{k! \ell! (n-k-\ell)!}\nn\\
    &\qd \times  I_{i+k}\big(1,(\alpha^*+\beta)\big) I_{j+\ell}\big(1,-\iu(\alpha^*-\beta)\big)
}

Now using Eq.\ \eqref{eq:Igen}, the displacement inner product can be directly written as:
\aln{
   \Bk{F_{\alpha}^{\star}(z)|F_{\beta}^{\star}(z)}_{\tx{SB}} &= \frac{n!}{\pi} \ec^{-(\Abs{\alpha}^2+\Abs{\beta}^2)/2} \sum_{i=0}^{n} \sum_{j=0}^{n-i} \sum_{k=0}^{n} \sum_{\ell=0}^{n-k} \frac{(-\iu)^j (-\alpha)^{n-i-j}}{i! j! (n-i-j)!} \frac{(\iu)^{\ell} (-\beta^*)^{n-k-\ell}}{k! \ell! (n-k-\ell)!}\nn\\
    &\qd \times \lb(\sqrt{\pi} \ec^{(\alpha^*+\beta)^2} \sum_{p=0}^{i+k} \gamma_{(i+k),p} (\alpha^*+\beta)^p\rb) \nn\\ 
    &\qd \times \lb(\sqrt{\pi} \ec^{-(\alpha^*-\beta)^2} \sum_{\qew=0}^{j+\ell} \gamma_{(j+\ell),\qew} \big(-\iu(\alpha^*-\beta)\big)^{\qew}\rb) \nn\\
   &= n!\ \ec^{-(\Abs{\alpha}^2+\Abs{\beta}^2)/2}\ \ec^{\alpha^*\beta}\ \sum_{i=0}^{n} \sum_{j=0}^{n-i} \sum_{k=0}^{n} \sum_{\ell=0}^{n-k} \sum_{p=0}^{i+k} \sum_{\qew=0}^{j+\ell} \frac{(-\iu)^j (-\alpha)^{n-i-j}}{i! j! (n-i-j)!} \nn\\
   &\qqd \times \frac{(\iu)^{\ell} (-\beta^*)^{n-k-\ell}}{k! \ell! (n-k-\ell)!} \ \gamma_{(i+k),p} \gamma_{(j+\ell),\qew} \big(\alpha^*+\beta\big)^p \big(-\iu(\alpha^*-\beta)\big)^\qew
}

which is the product of a Gaussian and a polynomial of degree $2n$ in $\alpha$ and $\beta$.

\subsection{Explicit examples of the displaced Fock state kernel}
\label{sec:KDExamples}

In this section we write down the explicit form of the first nine displacement kernels.

\begin{equation*}
\def\arraystretch{1.4}
\begin{array}{@{}cl@{}}
\toprule
    \tx{Initial Fock state},\ \Ket{n} &
        \multicolumn{1}{c}{\tx{Kernel function},\  %
        k(\bd{\alpha},\bd{\beta})} \\
\cmidrule(lr){1-2}
    \Ket{0} &
       \ec^{-\Abs{\bd{\alpha}-\bd{\beta}}^2} \\
    \Ket{1} &
        \ec^{-\Abs{\bd{\alpha}-\bd{\beta}}^2}\big(\Abs{\bd{\alpha}-\bd{\beta}}^2 - 1 \big)^2 \\
    \Ket{2} &
        \frac{\ec^{-\Abs{\bd{\alpha}-\bd{\beta}}^2}}{4}\big(2 - 4\Abs{\bd{\alpha}-\bd{\beta}}^2 + \Abs{\bd{\alpha}-\bd{\beta}}^4\big)^2 \\
    \Ket{3} &
        \frac{\ec^{-\Abs{\bd{\alpha}-\bd{\beta}}^2}}{36} \big(-6 + 18\Abs{\bd{\alpha}-\bd{\beta}}^2 - 9\Abs{\bd{\alpha}-\bd{\beta}}^4 + \Abs{\bd{\alpha}-\bd{\beta}}^6\big)^2 \\
    \Ket{4} &
        \frac{\ec^{-\Abs{\bd{\alpha}-\bd{\beta}}^2}}{576} \big(24 - 96\Abs{\bd{\alpha}-\bd{\beta}}^2 + 72\Abs{\bd{\alpha}-\bd{\beta}}^4 - 16\Abs{\bd{\alpha}-\bd{\beta}}^6 + \Abs{\bd{\alpha}-\bd{\beta}}^8\big)^2 \\
    \Ket{5} &
        \frac{\ec^{-\Abs{\bd{\alpha}-\bd{\beta}}^2}}{14400} \big(-120 + 600\Abs{\bd{\alpha}-\bd{\beta}}^2 - 600\Abs{\bd{\alpha}-\bd{\beta}}^4 + 200\Abs{\bd{\alpha}-\bd{\beta}}^6 \\
    &
        \qqqd - 25\Abs{\bd{\alpha}-\bd{\beta}}^8 +\Abs{\bd{\alpha}-\bd{\beta}}^{10}\big)^2 \\
    \Ket{6} &
        \frac{\ec^{-\Abs{\bd{\alpha}-\bd{\beta}}^2}}{518400} \big(720 - 4320\Abs{\bd{\alpha}-\bd{\beta}}^2 + 5400\Abs{\bd{\alpha}-\bd{\beta}}^4 - 2400\Abs{\bd{\alpha}-\bd{\beta}}^6 \\
    &
        \qqqd + 450\Abs{\bd{\alpha}-\bd{\beta}}^8 - 36\Abs{\bd{\alpha}-\bd{\beta}}^{10} + \Abs{\bd{\alpha}-\bd{\beta}}^{12}\big)^2 \\
    \Ket{7} &
        \frac{\ec^{-\Abs{\bd{\alpha}-\bd{\beta}}^2}}{25401600} \big(-5040 + 35280\Abs{\bd{\alpha}-\bd{\beta}}^2 - 52920\Abs{\bd{\alpha}-\bd{\beta}}^4 + 29400\Abs{\bd{\alpha}-\bd{\beta}}^6 \\
    &
        \qqqd  - 7350\Abs{\bd{\alpha}-\bd{\beta}}^8 + 882\Abs{\bd{\alpha}-\bd{\beta}}^{10}  - 49\Abs{\bd{\alpha}-\bd{\beta}}^{12} +\Abs{\bd{\alpha}-\bd{\beta}}^{14}\big)^2 \\
    \Ket{8} &
        \frac{\ec^{-\Abs{\bd{\alpha}-\bd{\beta}}^2}}{1625702400} \big(40320 - 322560\Abs{\bd{\alpha}-\bd{\beta}}^2 + 564480\Abs{\bd{\alpha}-\bd{\beta}}^4 - 376320\Abs{\bd{\alpha}-\bd{\beta}}^6 \\
    &
        \qqqd + 117600\Abs{\bd{\alpha}-\bd{\beta}}^8 - 18816 \Abs{\bd{\alpha}-\bd{\beta}}^{10} + 1568 \Abs{\bd{\alpha}-\bd{\beta}}^{12} \\
    &
        \qqqd - 64 \Abs{\bd{\alpha}-\bd{\beta}}^{14} + \Abs{\bd{\alpha}-\bd{\beta}}^{16}\big)^2\\
\bottomrule
\end{array}
\end{equation*}

\subsection{Showing the displaced Fock state kernel is translation invariant}
\label{sec:KDShift}

A translation invariant kernel has the property that 
\eqn{
    k(\bd{\alpha}+\bd{h},\bd{\beta}+\bd{h}) = k(\bd{\alpha},\bd{\beta}).
}
For the case of the displaced Fock state kernel, the vector $\bd{h}=(h_1,h_2)^\trps\in\bb{R}^2$, and we define $h=h_1+\iu h_2$. Now the shifted kernel is
\eqn{
	k(\bd{\alpha}+\bd{h},\bd{\beta}+\bd{h}) = \Abs{\Bk{F_{\alpha+h}^{\star}(z)|F_{\beta+h}^{\star}(z)}_{\tx{SB}}}^2
}
where
\aln{
    &\Bk{F_{\alpha+h}^{\star}(z)|F_{\beta+h}^{\star}(z)}_{\tx{SB}} \nn\\
    &\qd = \frac{1}{\pi} \frac{\ec^{-(\Abs{\alpha+h}^2+\Abs{\beta+h}^2)/2}}{n!} \int_{z\in\bb{C}} \dd^2z\ \ec^{-[\Abs{z}^2-(\alpha^*+h^*)z^*-(\beta+h)z]}(z^*-\alpha-h)^n(z-\beta^*-h^*)^n.
}
Change variables $z\to z+h^*$, so the inner product is now
\aln{
    \Bk{F_{\alpha+h}^{\star}(z)|F_{\beta+h}^{\star}(z)}_{\tx{SB}} &= \frac{1}{\pi} \frac{\ec^{-(\Abs{\alpha+h}^2+\Abs{\beta+h}^2)/2}}{n!} \int_{z\in\bb{C}} \dd^2z\ \ec^{-\Abs{z+h^*}^2+(\alpha^*+h^*)(z^*+h)+(\beta+h)(z+h^*)} \nn\\
    &= \frac{1}{\pi} \frac{\ec^{-(\Abs{\alpha}^2+\Abs{\beta}^2)/2}\ \ec^{[h(\alpha^*-\beta^*) - h^*(\alpha-\beta)]/2}}{n!} \nn\\
    &\qd \times \int_{z\in\bb{C}} \dd^2z\ \ec^{-(\Abs{z}-\alpha^*z^*-\beta z)} (z^*-\alpha)^n (z-\beta^*)^n \nn\\
    &= \ec^{\iu\Im[h(\alpha^*-\beta^*)]} \Bk{F_{\alpha}^{\star}(z)|F_{\beta}^{\star}(z)}_{\tx{SB}}.
    \label{eq:Dkerneltrans}
}

Therefore:
\aln{
	k(\bd{\alpha}+\bd{h},\bd{\beta}+\bd{h}) &= \Abs{\Bk{F_{\alpha+h}^{\star}(z)|F_{\beta+h}^{\star}(z)}_{\tx{SB}}}^2 \nn\\
	&= \lb(\ec^{\iu\Im[h(\alpha^*-\beta^*)]} \Bk{F_{\alpha}^{\star}(z)|F_{\beta}^{\star}(z)}_{\tx{SB}}\rb) \lb(\ec^{-\iu\Im[h(\alpha^*-\beta^*)]} \Bk{F_{\alpha}^{\star}(z)|F_{\beta}^{\star}(z)}_{\tx{SB}}^*\rb) \nn\\
	&= \Abs{\Bk{F_{\alpha}^{\star}(z)|F_{\beta}^{\star}(z)}_{\tx{SB}}}^2 \nn\\
	&= k(\bd{\alpha},\bd{\beta})
}
and the displacement kernel is translation invariant.

\subsection{Showing the displaced Fock state kernel is rotation invariant}
\label{sec:KDRotation}

A rotation invariant kernel has the property that
\eqn{
    k\big(\mc{R}(\theta)\bd{\alpha},\mc{R}(\theta)\bd{\beta}\big) = k(\bd{\alpha},\bd{\beta}).
}
For the case of the displaced Fock state kernel, the rotation matrix is
\eqn{
	\mc{R}(\theta) = \pmx{
		\cos(\theta) & -\sin(\theta) \\
		\sin(\theta)  & \cos(\theta)
	}.
}
Under this transformation, the complex number $\alpha\to\ec^{\iu\theta}\alpha$, so the rotated kernel is
\eqn{
    k\big(\mc{R}(\theta)\bd{\alpha},\mc{R}(\theta)\bd{\beta}\big) = \Abs{\Bk{F_{\exp(\iu\theta)\alpha}^{\star}(z)|F_{\exp(\iu\theta)\beta}^{\star}(z)}_{\tx{SB}}}^2,
}
where
\aln{
    \Bk{F_{\exp(\iu\theta)\alpha}^{\star}(z)|F_{\exp(\iu\theta)\beta}^{\star}(z)}_{\tx{SB}} &= \frac{1}{\pi} \frac{\ec^{-(\Abs{\alpha}^2+\Abs{\beta}^2)/2}}{n!} \int_{z\in\bb{C}} \dd^2z\ \ec^{-[\Abs{z}^2-\exp(-\iu\theta)\alpha^*z^*-\exp(\iu\theta)\beta z]} \nn\\
    &\qd \times \lb(z^*-\ec^{\iu\theta}\alpha\rb)^n \lb(z+\ec^{-\iu\theta}\beta^*\rb)^n.
}
Change variables $z\to \ec^{-\iu\theta}z$, so the inner product is now
\aln{
    \Bk{F_{\exp(\iu\theta)\alpha}^{\star}(z)|F_{\exp(\iu\theta)\beta}^{\star}(z)}_{\tx{SB}} &= \frac{1}{\pi} \frac{\ec^{-(\Abs{\alpha}^2+\Abs{\beta}^2)/2}}{n!} \nn\\
    &\qd \times \int_{z\in\bb{C}} \dd^2z\ \ec^{-[\Abs{z}^2-\exp(-\iu\theta)\alpha^*\exp(\iu\theta)z^*-\exp(\iu\theta)\beta \exp(-\iu\theta)z]} \nn\\
    &\qqd \times \lb(\ec^{\iu\theta}z^*-\ec^{\iu\theta}\alpha\rb)^n \lb(\ec^{-\iu\theta}z+\ec^{-\iu\theta}\beta^*\rb)^n \nn\\
    &= \frac{1}{\pi} \frac{\ec^{-(\Abs{\alpha}^2+\Abs{\beta}^2)/2}}{n!} \int_{z\in\bb{C}} \dd^2z\ \ec^{-(\Abs{z}^2-\alpha^*z^*-\beta z)}\nn\\
    &\qd \times \ec^{n\iu\theta}\ec^{-n\iu\theta} (z^*-\alpha)^n (z+\beta^*)^n \nn\\
    & = \frac{1}{\pi} \frac{\ec^{-(\Abs{\alpha}^2+\Abs{\beta}^2)/2}}{n!} \int_{z\in\bb{C}} \dd^2z\ \ec^{-(\Abs{z}^2-\alpha^*z^*-\beta z)} \nn\\
    &\qd \times (z^*-\alpha)^n (z+\beta^*)^n \nn\\
    &= \Bk{F_{\alpha}^{\star}(z)|F_{\beta}^{\star}(z)}_{\tx{SB}}
}
and
the displaced Fock state kernel is rotational invariant.

\subsection{Showing the displaced Fock state kernel is a radial kernel}
\label{sec:KDRadial}

From these translation and rotation invariance of the displaced Fock state kernel, can also show that
\aln{
    k\big(\mc{R}(\theta)(\bd{\alpha}-\bd{\beta})\big) 
    &= k\big(\mc{R}(\theta)\bd{\alpha}-\mc{R}(\theta)\bd{\beta}\big) \nn\\
    &= k\big(\mc{R}(\theta)\bd{\alpha},\mc{R}(\theta)\bd{\beta}\big)\nn\\
    &= k(\bd{\alpha},\bd{\beta}) \nn\\
    &= k(\bd{\alpha}-\bd{\beta}),
}
and therefore
\eqn{
    k(\bd{\alpha},\bd{\beta}) = k(\Abs{\bd{\alpha}-\bd{\beta}})
}
it is a radial kernel. 

Furthermore, the polynomial $P(\alpha,\beta)$ is a polynomial of $\alpha_1$, $\alpha_2$, $\beta_1$ and $\beta_2$, but, there is no way of constructing 
\eqn{
	\Abs{\bd{s}} = \sqrt{(\alpha_1-\beta_1)^2+(\alpha_2-\beta_2)^2}
} 
out of such a polynomial. However there is a way of constructing
\eqn{
	\Abs{\bd{s}}^2 = (\alpha_1-\beta_1)^2+(\alpha_2-\beta_2)^2 = \alpha_1^2+\alpha_2^2+\beta_1^2+\beta_2^2-2\alpha_1\beta_1-2\alpha_2\beta_2
}
and therefore
\eqn{
    k(\bd{\alpha},\bd{\beta}) = k(\Abs{\bd{\alpha}-\bd{\beta}}^2)
}

\subsection{The Fourier transform of the displaced Fock state kernel}
\label{sec:KDFT}

Now define $\bd{s}\ce\bd{\alpha}-\bd{\beta}$, and use the facts that the displaced Fock state kernel is
\eqn{
   k(\bd{\alpha},\bd{\beta}) = k(\Abs{\bd{s}}^2)
}
and is the product of a Gaussian and a polynomial of degree $4n$ in $\bd{s}$, to write is as
\eqn{
    k\lb(\Abs{\bd{s}}\rb) = \ec^{-\Abs{\bd{s}}^2} \sum_{j=0}^{2n} a_{2j} \Abs{\bd{s}}^{2j}
}
for an appropriate choice of $a_j\in\bb{R}$. 

The two-dimensional Fourier transform can be easily calculated as
\aln{
    \frac{1}{2\pi} \int \ddd^2 s\ \ \ec^{\iu\bd{s}\cdot\bd{\omega}} k(\bd{s}) &=  \frac{1}{2\pi} \sum_{j=0}^{2n} a_{2j} \intallr \ddd s\ \int_{0}^{2\pi} \ddd\theta\ s \ec^{\iu s\omega\cos(\theta)} \ec^{-s^2} s^{2j} \nn\\
    &= \frac{1}{2\pi} \sum_{j=0}^{2n} a_{2j} \intallr \ddd s\ \ec^{-s^2} s^{2j+1} J_0(\omega s) \nn\\
    &= \frac{1}{2\pi} \sum_{j=0}^{2n} a_{2j} \lb[\frac{j!}{2}\ \hypg{1}{F}{1}\lb(1+j;\ 1;\ -\frac{\omega^2}{4}\rb)\rb]
}
which can be simplified further using the hypergeometric identity in appendix \ref{sec:hypg}
\aln{
    \frac{1}{2\pi} \int \ddd^2 s\ \ \ec^{\iu\bd{s}\cdot\bd{\omega}} k(\bd{s}) &= \frac{1}{4\pi} \sum_{j=0}^{2n} j!\ a_{2j}\ \ec^{-\omega^2/4} \sum_{\ell=0}^{j} {j \choose \ell} \frac{1}{(1)_\ell} \lb(-\frac{\omega^2}{4}\rb)^\ell \nn\\
    &= \frac{\ec^{-\omega^2/4}}{4\pi} \sum_{j=0}^{2n} (j!)^2 a_{2j}\  \sum_{\ell=0}^{j} \frac{(-1)^\ell}{(\ell!)^2(j-\ell)!} \lb(\frac{\omega}{2}\rb)^{2\ell}
}
which is also the product of a Gaussian and a polynomial of degree $4n$ in $\omega$.

\subsection{Showing the displaced Fock state kernel integrates to $\pi$}
\label{sec:KDintegral}

In this section, we calculate the displaced Fock state kernel from the operator definition:
\eqn{
    \hat{D}(\alpha) = \ec^{\alpha\oad - \alpha^*\oa} = \ec^{-\Abs{\alpha}^2/2}\ec^{\alpha\oad}\ec^{-\alpha^*\oa}.
}

Starting from 
\aln{
    \ec^{\pm \alpha^*\oa}\Ket{n} &= \sum_{j=0}^{\infi} \frac{(\pm \alpha^*)^j\oa^j}{j!}\Ket{n} \nn\\
    &= \Ket{n} + \sum_{j=1}^{\infi} \frac{(\pm \alpha^*)^j\oa^j}{j!}\Ket{n} \nn\\
    &=  \Ket{n} + \sum_{j=1}^{n} \frac{(\pm \alpha^*)^j}{j!} \prod_{\ell=1}^{j} \sqrt{n-\ell+1} \Ket{n-j}.
}
we calculate the inner product as
\aln{
    \Bk{n|\hat{D}^{\ct}(\alpha)\hat{D}(\beta)|n} &= \ec^{(-\alpha\beta^*+\alpha^*\beta)/2}\Bk{n|\hat{D}(\beta-\alpha)|n} \nn\\
    &= \ec^{-\Abs{\beta-\alpha}^2/2} \ec^{\iu\Im(\alpha^*\beta)} \Bk{n|\ec^{(\beta-\alpha)\oad}\ec^{-(\beta-\alpha)^*\oa}|n} \nn\\
    &= \ec^{-\Abs{(\beta-\alpha)}^2/2} \ec^{\iu\Im(\alpha^*\beta)} \lb(\ec^{(\beta-\alpha)^*\oa}\Ket{n}\rb)^\ct \lb(\ec^{-(\beta-\alpha)^*\oa}\Ket{n}\rb) \nn\\
    &= \ec^{-\Abs{\beta-\alpha}^2/2} \ec^{\iu\Im(\alpha^*\beta)} \lb(\Bra{n} + \sum_{i=1}^{n} \frac{(\beta-\alpha)^i}{i!} \prod_{k=1}^{i} \sqrt{n-k+1} \Bra{n-i}\rb) \nn\\
	&\qqd \times \Bigg(\Ket{n} + \sum_{j=1}^{n} \frac{\big(-(\beta-\alpha)^*\big)^j}{j!} \prod_{\ell=1}^{j} \sqrt{n-\ell+1} \Ket{n-j}\Bigg) \nn\\
    &= \ec^{-\Abs{\beta-\alpha}^2/2} \ec^{\iu\Im(\alpha^*\beta)} \lb(1+\sum_{j=1}^{n} \frac{(\beta-\alpha)^j\big(-(\beta-\alpha)^*\big)^j}{(j!)^2} \lb(\sqrt{n}\sqrt{n-1}\dotsc\sqrt{n-j+1}\rb)^2\rb) \nn\\
    &= \ec^{-\Abs{\beta-\alpha}^2/2} \ec^{\iu\Im(\alpha^*\beta)} \sum_{j=0}^{n} \frac{(-1)^j\Abs{\beta-\alpha}^{2j}}{(j!)^2} \frac{n!}{(n-j)!} \nn\\
    &= \ec^{-\Abs{\beta-\alpha}^2/2} \ec^{\iu\Im(\alpha^*\beta)} \sum_{j=0}^{n} {n \choose j} \frac{(-1)^j\Abs{\beta-\alpha}^{2j}}{j!}
}
and the displaced Fock state kernel
\eqn{
	k(\bd{\alpha},\bd{\beta}) = \Abs{\Bk{n|\hat{D}^{\ct}(\alpha)\hat{D}(\beta)|n}}^2 = \ec^{-\Abs{\beta-\alpha}^2} \lb(\sum_{j=0}^{n} {n \choose j} \frac{(-1)^j\Abs{\beta-\alpha}^{2j}}{j!} \rb)^2.
}
Note that this is a much simpler form than that which we found Eq.\ \eqref{eq:Dkernel}, indicating that there may be a way to significantly simplify the general multi-mode kernel Eq.\ \eqref{eq:MMkernel}. We leave this exploration for future work.

Now define $\bd{s}\ce\bd{\alpha}-\bd{\beta} \in \bb{R}^2$ and integrate the kernel over all $\bd{s}$
\aln{
    \int \dd^2s\ k(\Abs{\bd{s}}) &= \intallr \dd s\ s \int_{0}^{2\pi} \dd \theta\ \ec^{-s^2} \lb(\sum_{j=0}^{n} {n \choose j} \frac{(-1)^j s^{2j}}{j!}\rb)^2 \nn\\
    &= 2\pi \sum_{j=0}^{n} \sum_{k=0}^{n} {n \choose j} {n \choose k} \frac{(-1)^j}{j!} \frac{(-1)^k}{k!} \intallr \dd s\ \ec^{-s^2} s^{2(j+k)+1} \nn\\
    &= 2\pi \sum_{j=0}^{n} \sum_{k=0}^{n} {n \choose j} {n \choose k} \frac{(-1)^{j+k}}{j!k!} \lb(\frac{\Gamma(1+j+k)}{2}\rb) \nn\\
    &= \pi \sum_{j=0}^{n} \sum_{k=0}^{n} (-1)^{j+k} {n \choose j} {n \choose k} {j+k \choose k}.
}

The binomial coefficients can be further simplified by using the following properties for $k,n\in\bb{N}_0$, $m\in\bb{Z}$, and $x,y\in\bb{R}$ \cite{wolfram:binomial}:
\subeqn{
\aln{
    {x \choose k} &= \frac{x^{\underline{k}}}{k!} \label{eq:binomdef}\\
    {m \choose k} &= (-1)^k {-m+k-1 \choose k} \label{eq:binomtop}\\
    {n \choose k} &= (-1)^{n-k} {-k-1 \choose n-k} \quad k\le n \label{eq:binomswitch}\\
    {x+y \choose n} &= \sum_{k=0}^{n} {x \choose k} {y \choose n-k} \label{eq:binomCV}
}
}
where
\eqn{
	x^{\underline{k}} = x(x-1)(x-2)\dotsc(x-k+1)
}
is the falling factorial.

With these properties, the integral becomes:
\aln{
    \int \dd^2s\ k(\Abs{\bd{s}}) &= \pi \sum_{j=0}^{n} \sum_{k=0}^{n} (-1)^{j+k} {n \choose j} {n \choose k} {j+k \choose k} \nn\\
    &= \pi \sum_{j=0}^{n} \sum_{k=0}^{n} (-1)^{j+k} {n \choose j} {n \choose k} (-1)^{j+k-k} {-(k+1) \choose j} &(\tx{Eq.\ \eqref{eq:binomswitch}}) \nn\\
    &= \pi \sum_{k=0}^{n} (-1)^k {n \choose k} \sum_{j=0}^{n} {n \choose n-j} {-(k+1) \choose j} \nn\\
    &= \pi \sum_{k=0}^{n} (-1)^k {n \choose k} {n-(k+1) \choose n} &(\tx{Eq.\ \eqref{eq:binomCV}}) \nn\\
    &= \pi \sum_{k=0}^{n} (-1)^k {n \choose k} (-1)^{n} {k \choose n} &(\tx{Eq.\ \eqref{eq:binomtop}}) \nn\\
    &= \pi \sum_{k=0}^{n} (-1)^{n+k} {n \choose k} \frac{k^{\underline{n}}}{n!} &(\tx{Eq.\ \eqref{eq:binomdef}}).
}

Since $0\le k \le n$, the falling factorial becomes
\eqn{
    k^{\underline{n}} = k(k-1)\dotsc(k-k)\dotsc(k-n+1)
}
which is zero unless $k=n$, so
\aln{
    \int \dd^2s\ k(\Abs{\bd{s}}) &= \pi \sum_{k=0}^{n} (-1)^{n+k} {n \choose k} \frac{k(k-1)\dotsc(k-n+1)}{n!} \delta_{k,n} \nn\\
    &= \pi (-1)^{2n} {n \choose n} \frac{n!}{n!} \nn\\
    &= \pi
}
for all $n\in\bb{N}_0$.

\section{Calculation of the qudit kernel}
\label{sec:Quditcalc}

In this section we show the details of the calculation of the qudit kernel (Eq.\eqref{eq:QuditKernel}),

In the function representation, the inner product of two qudits is
\aln{
    \Bk{F^{\star}_{\txx_1}(z)|F^{\star}_{\txx_2}(z)}_{\tx{SB}} &= \frac{1}{\pi} \int_{z\in\bb{C}} \ddd z^2\ \ec^{-\Abs{z}^2} \vast(\sum_{i=0}^{d-1} \ms{n}_{d,i}^*\alpha_i(\txx_1)^* (z^*)^i \vast)\lb(\sum_{j=0}^{d-1} \ms{n}_{d,j}\alpha_j(\txx_2) z^j\rb) \nn\\
    &= \frac{1}{\pi} \sum_{i=0}^{d-1} \sum_{j=0}^{d-1} \ms{n}_{d,i}^* \ms{n}_{d,j} \alpha_i(\txx_1)^* \alpha_j(\txx_2) \intallr \ddd r\ r \int_{0}^{2\pi} \ddd\theta\ \ec^{-r^2} \lb(r\ec^{-\iu\theta}\rb)^i \lb(r\ec^{\iu\theta}\rb)^j \nn\\
    &= \frac{1}{\pi} \sum_{i=0}^{d-1} \sum_{j=0}^{d-1} \ms{n}_{d,i}^* \ms{n}_{d,j} \alpha_i(\txx_1)^* \alpha_j(\txx_2) \lb(\intallr \ddd r\ \ec^{-r^2} r^{i+j+1}\rb) \lb(\int_{0}^{2\pi} \ddd\theta\ \ec^{\iu(j-i)\theta}\rb) \nn\\
    &= \frac{1}{\pi} \sum_{i=0}^{d-1} \sum_{j=0}^{d-1} \ms{n}_{d,i}^* \ms{n}_{d,j} \alpha_i(\txx_1)^* \alpha_j(\txx_2) \Bigg(\frac{1}{2} \Gamma\lb(1+\frac{i+j}{2}\rb)\Bigg) \Big(2\pi \delta_{i,j}\Big) \nn\\
    &= \sum_{j=0}^{d-1} j! \Abs{\ms{n}_{d,j}}^2 \alpha_j(\txx_1)^* \alpha_j(\txx_2).
    \label{eq:QuditKernelpolar}
}

\subsection{Calculating the value of $\ms{n}_{d,j}$}
\label{sec:QuditNormalization}
Recall, that in the case when $\txx_1=\txx_2$, $\Ket{F^{\star}_2(z)} = \Ket{F^{\star}_1(z)}$,so 
\aln{
    \Bk{F^{\star}_1(z)|F^{\star}_1(z)}_{\tx{SB}} = \sum_{j=0}^{d-1} j! \Abs{\ms{n}_{d,j}}^2 \Abs{\alpha_j(\txx_1)}^2 = \sum_{j=0}^{d-1} \Abs{\alpha_j(\txx_1)}^2 = 1,
}
where the second equality results from the definition of the qudit stellar function (Eq.\ \eqref{eq:QuditNormailzation}).
Therefore,
\eqn{
    \Abs{\ms{n}_{d,j}}^2  = \frac{1}{j!}
}
so we choose
\eqn{
    \ms{n}_{d,j} \ce \frac{1}{\sqrt{j!}}.
    \label{eq:QuditnDef}
}

\subsection{Calculation of the qudit kernel from the general multi-mode kernel}
\label{sec:Quditcalc2}

In this section, we calculate the qudit kernel from the general multi-mode kernel (Eq.\ \eqref{eq:MMkernel}). 

First we note that when $x=0$
\eqn{
    \sum_{j=0}^{n} \alpha_j x^j = \sum_{j=0}^{n} \alpha_j \delta_{j,0} = \alpha_0.
}

Since a qudit can always be represented as a function of a single complex variable, we set $m=1$. Additionally from Eq.\ \eqref{eq:Quditstate} we find that
\alns{
    a_{0,j} &= 1 \\
    b_{0,j} &= 0 \\
    d_{0,j,k} &= 0 \\
    C(\txx_1) &= C(\txx_2) = 0
}
and by the recursion relations (Eq.\ \eqref{eq:Recusion1})
\alns{
    a_{i,j} &= 1 \\
    b_{i,j} &= 0 \\
    d_{i,j,k} &= 0.
}

Now the inner product becomes
\aln{
    \Bk{F^{\star}_{\txx_1}(z)|F^{\star}_{\txx_2}(z)}_{\tx{SB}} &= \sum_{i=0}^{n} \sum_{j=0}^{n} \beta_i(\txx_1)^* \beta_j(\txx_2) \sum_{p=0}^{i} \sum_{\qew=0}^{j} g(i,j,p,\qew) \sum_{s_1=0}^{r_{0,1}} \gamma_{r_{0,1},s_1} \nn\\
    &\qd \times \sum_{t_1=0}^{s_1} \frac{s_1!}{(s_1-t_1)!} \delta_{t_1,s_1} \frac{\delta_{t_1,0}}{t_1!} \sum_{s_2=0}^{r_{1,2}} \gamma_{r_{1,2},s_2}\ \delta_{s_2,0} \nn\\
    &= \sum_{i=0}^{n} \sum_{j=0}^{n} \beta_i(\txx_1)^* \beta_j(\txx_2) \sum_{p=0}^{i} \sum_{\qew=0}^{j} g(i,j,p,\qew) \sum_{s_1=0}^{r_{0,1}} \gamma_{r_{0,1},s_1} \nn\\
    &\qqd \times \sum_{t_1=0}^{s_1} \frac{s_1!}{(s_1-t_1)!} \delta_{t_1,s_1} \frac{\delta_{t_1,0}}{t_1!}  \sum_{s_2=0}^{r_{0,2}+t_1} \gamma_{(r_{0,2}+t_1),s_2}\ \delta_{s_2,0} \nn\\
    &= \sum_{i=0}^{n} \sum_{j=0}^{n} \beta_i(\txx_1)^* \beta_j(\txx_2) \sum_{p=0}^{i} \sum_{\qew=0}^{j} g(i,j,p,\qew) \gamma_{r_{0,1},0} \gamma_{r_{0,2},0} \nn\\
    &= \sum_{i=0}^{n} \sum_{j=0}^{n} \beta_i(\txx_1)^* \beta_j(\txx_2) \sum_{p=0}^{i} \sum_{\qew=0}^{j} {i\choose p} {j\choose\qew} (-\iu)^{p} (\iu)^{\qew} \gamma_{(i+j-p-\qew),0} \gamma_{(p+\qew),0}
}
where in the second line, we substitute in the recursion relation for $r$ (Eq.\ \eqref{eq:Recusion2}) and in the fourth line we substitute in the values of $g$ and $r_{0,j}$ (Eq.\ \eqref{eq:seeds}).

From Eq.\ \eqref{eq:gammagen},
\aln{
    \gamma_{r,0} &= \begin{dcases}
        \frac{1}{2^r} \frac{r!}{(r/2)!}, & r\ \tx{even}\\
        0,& \tx{otherwise}
    \end{dcases} \nn\\
    &= \frac{1}{\sqrt{\pi}} \frac{1+(-1)^r}{2} \Gamma\lb(\frac{1}{2}+\frac{r}{2}\rb)
}
and so
\aln{
    \Bk{F^{\star}_{\txx_1}(z)|F^{\star}_{\txx_2}(z)}_{\tx{SB}} &= \sum_{i=0}^{n} \sum_{j=0}^{n} \beta_i(\txx_1)^* \beta_j(\txx_2) \sum_{p=0}^{i} \sum_{\qew=0}^{j} {i \choose p} {j \choose \qew} (-\iu)^p (\iu)^\qew \nn\\
    &\qqd \times \Bigg(\frac{1}{\sqrt{\pi}} \frac{1+(-1)^{i+j-p-\qew}}{2} \Gamma\lb(\frac{1}{2}+\frac{i+j-p-\qew}{2}\rb)\Bigg) \nn\\
    &\qqd \times \Bigg(\frac{1}{\sqrt{\pi}} \frac{1+(-1)^{p+\qew}}{2} \Gamma\lb(\frac{1}{2}+\frac{p+\qew}{2}\rb)\Bigg) \nn\\
    &= \frac{1}{\pi} \sum_{i=0}^{n} \sum_{j=0}^{n} \beta_i(\txx_1)^* \beta_j(\txx_2) \sum_{p=0}^{i} \sum_{\qew=0}^{j} {i \choose p} {j \choose \qew} (-\iu)^p (\iu)^\qew \nn\\
    &\qqd \times \lb(\intall \ddd x\ \ec^{-x^2}x^{i+j-p-\qew}\rb) \lb(\intall \ddd y\ \ec^{-y^2}y^{p+\qew}\rb) \nn\\
    &= \frac{1}{\pi} \intall \ddd x\ \intall \ddd y\ \ec^{-(x^2+y^2)} \sum_{i=0}^{n} \sum_{j=0}^{n} \beta_i(\txx_1)^* \beta_j(\txx_2) \nn\\
    &\qqd \times \Bigg[{i\choose p} x^{i-p} (-\iu y)^p \Bigg] \Bigg[{j\choose\qew} x^{j-\qew} (\iu y)^{\qew}\Bigg] \nn\\
    &= \frac{1}{\pi} \intall \ddd x\ \intall \ddd y\ \ec^{-(x^2+y^2)} \sum_{i=0}^{n} \sum_{j=0}^{n} \beta_i(\txx_1)^* \beta_j(\txx_2) (x-\iu y)^i (x+\iu y)^j \nn\\
    &= \frac{1}{\pi} \int_{z\in\bb{C}} \ddd^2 z\ \ec^{-\Abs{z}^2} \vast(\sum_{i=0}^{n} \beta_i(\txx_1)^* (z^*)^i\vast) \lb(\sum_{j=0}^{n} \beta_j(\txx_2) z^j\rb)
    \label{eq:QuditKernelGen}
}
where in the second line, we use the fact that for $r\in\bb{N}_0$,
\eqn{
    \intall \ddd x\ \ec^{-x^2} x^r = \frac{1+(-1)^r}{2} \Gamma\lb(\frac{1}{2}+\frac{r}{2}\rb).
}
Clearly, this matches Eq.\ \eqref{eq:QuditKernelpolar} for $n=d-1$ and $\beta_j = \ms{n}_{d,j} \alpha_j = \alpha_j/\sqrt{j!}$.

\end{document}